\documentclass[%
 reprint,
superscriptaddress,
 amsmath,amssymb,
 aps,
]{revtex4-1}

\usepackage{graphicx}
\usepackage{dcolumn}
\usepackage{bm}

\usepackage{here}

\newcommand{\mlm}[1]{\textcolor{black}{#1}}

\newcommand{\ty}[1]{\textcolor{black}{#1}}

\newcommand{\blue}[1]{\textcolor{black}{#1}}

\usepackage{color}
\usepackage{comment}
\begin{document}

\preprint{APS/123-QED}

\title{Non-monotonic fluidization generated by fluctuating edge tensions in confluent tissues}

\author{Takaki Yamamoto}
\email{takaki.yamamoto@riken.jp}
\affiliation{Laboratory for Physical Biology, RIKEN Center for Biosystems Dynamics Research, Kobe 650-0047, Japan}
\affiliation{Nonequilibrium Physics of Living Matter RIKEN Hakubi Research Team, RIKEN Center for Biosystems Dynamics Research, 2-2-3 Minatojima-minamimachi, Chuo-ku, Kobe 650-0047, Japan}
 
\author{Daniel M. Sussman}%
\affiliation{Department of Physics, Emory University, Atlanta, GA, USA}%
\author{Tatsuo Shibata}
\affiliation{Laboratory for Physical Biology, RIKEN Center for Biosystems Dynamics Research, Kobe 650-0047, Japan}

\author{M. Lisa Manning}%
 \email{memanning@gmail.com}
\affiliation{Department of Physics, Syracuse University, Syracuse, New York 13244, USA
}%
\affiliation{BioInspired Institute, Syracuse University, Syracuse, New York 13244, USA}%

\date{\today}

\begin{abstract}
{In development and homeostasis, multi-cellular systems exhibit spatial and temporal heterogeneity in their biochemical and mechanical properties. Nevertheless, it remains unclear how spatiotemporally heterogeneous forces affect the dynamical and mechanical properties of confluent tissue. To address this question, we study the dynamical behavior of the two-dimensional cellular vertex model for epithelial monolayers in the presence of fluctuating cell-cell interfacial tensions, which is a biologically relevant source of mechanical spatiotemporal heterogeneity. 
In particular, we investigate the effects of the amplitude and persistence time of fluctuating tension on the tissue dynamics.  We unexpectedly find that the long-time diffusion constant describing cell rearrangements depends non-monotonically on the persistence time, while it increases monotonically as the amplitude increases. 
Our analysis indicates that \mlm{at low and intermediate persistence times tension fluctuations drive motion of vertices and promote cell rearrangements, while at the highest persistence times the tension in the network evolves so slowly that rearrangements become rare.}}
\end{abstract}
\maketitle
\section{Introduction}
Spatiotemporal heterogeneity plays important roles in various biological processes~\cite{arigaNonequilibriumEnergeticsMolecular2018,arigaExperimentalTheoreticalEnergetics2020,hayashiApplicationFluctuationTheorem2018,yanagidaSingleMoleculeMeasurements2008,toyabeThermodynamicEfficiencyMechanochemical2011,fletcherCellMechanicsCytoskeleton2010,yamadaMechanisms3DCell2019,kleinUniversalPatternsStem2011a,barresiDevelopmentalBiology2019,mammotoMechanicalControlTissue2010a,shawkyTissueMechanicsAdhesion2015,petridouTissueRheologyEmbryonic2019}.
At the molecular scale, molecular motors such as kinesin~\cite{arigaNonequilibriumEnergeticsMolecular2018,arigaExperimentalTheoreticalEnergetics2020,hayashiApplicationFluctuationTheorem2018}, myosin~\cite{yanagidaSingleMoleculeMeasurements2008} and F1-ATPase~\cite{toyabeThermodynamicEfficiencyMechanochemical2011,hayashiApplicationFluctuationTheorem2018} utilize thermal temporal fluctuations to function. 
At \blue{the} scale of single cells, structures such as the cytoskeleton and focal adhesions spatially self-organize to execute necessary cellular functions~\cite{fletcherCellMechanicsCytoskeleton2010,yamadaMechanisms3DCell2019}. 
Finally, at the multi-cellular scale, spatiotemporal heterogeneity of gene expression and downstream cell differentiation are necessary for tissue homeostasis~\cite{kleinUniversalPatternsStem2011a} and proper development~\cite{barresiDevelopmentalBiology2019}.  
Cooperatively with this biochemical heterogeneity, multi-cellular systems control mechanical properties and cell motility to establish and maintain structures such as compartments and organs, and drive morphogenetic processes such as gastrulation and invagination~\cite{mammotoMechanicalControlTissue2010a,shawkyTissueMechanicsAdhesion2015,petridouTissueRheologyEmbryonic2019}. 
Therefore, it is essential to understand how spatiotemporally heterogeneous forces in multi-cellular systems affect the dynamical and mechanical properties of the tissue. 

Work over the past decade has suggested that the physics of jamming and glasses is a good starting point for understanding the mechanics and dynamics of multicellular tissues. Experiments have shown that dense biological tissues undergo solid-to-fluid transitions~\cite{parkUnjammingCellShape2015,fodorSpatialFluctuationsVertices2018a,mongeraFluidtosolidJammingTransition2018,angeliniGlasslikeDynamicsCollective2011b,wangAnisotropyLinksCell2020}, and near such transitions many systems, including Madin-Darby canine kidney (MDCK) cells~\cite{angeliniGlasslikeDynamicsCollective2011b} and primary human bronchial epithelial cells (HBECs)~\cite{parkUnjammingCellShape2015}, exhibit heterogeneous dynamics that are a hallmark of glassy dynamics. Recent work {\it in vivo} suggests that zebrafish use a spatial gradient in the fluid-to-solid transition to help drive body axis elongation~\cite{mongeraFluidtosolidJammingTransition2018}.
Also, theoretical studies have elucidated such glassy behaviors using mathematical models of confluent tissues such as the cellular vertex model (CVM)~\cite{sussmanAnomalousGlassyDynamics2018a,krajncSolidFluidTransition2020,kimEmbryonicTissuesActive2020}, the voronoi model (VM)~\cite{biMotilityDrivenGlassJamming2016a,sussmanAnomalousGlassyDynamics2018a} and the cellular Potts models~\cite{chiangGlassTransitionsCellular2016}. For instance, a VM study by Bi {\it et al.} reported that fluctuations induced by self-propulsion of the cells works in concert with cell mechanics to induce solid-to-fluid transitions~\cite{biMotilityDrivenGlassJamming2016a}. 
Some of us demonstrated anomalous glassy behavior in 2D confluent tissue driven by Brownian fluctuations in both CVM and VM ~\cite{sussmanAnomalousGlassyDynamics2018a}. Very recent work, initiated independently and concurrently with the work reported here, studied the effect of fluctuating tensions on confluent~\cite{krajncSolidFluidTransition2020} and non-confluent~\cite{kimEmbryonicTissuesActive2020} \blue{CVMs}. 
In general, all of these models agree that increasing either the magnitude of the fluctuating forces, or the persistence of such forces, can drive systems from the solid phase to the fluid phase.

In contrast, Yan {\it et al.} report on a mechanism that can drive a confluent tissue in the other direction, from a fluid state to a solid state~\cite{yanMulticellularRosettesDrive2019}. While all \blue{VMs}  and most \blue{CVMs} restrict allowable topologies to 3-fold coordinated vertices, Yan and collaborators demonstrate that introducing rosette structures, which are $n$-fold vertices ($n>3$), imposes topological constraints on the network of the CVM that can rigidify the tissue~\cite{yanMulticellularRosettesDrive2019} in static calculations. Recent work has studied the effect of explicit pinning of rosette structures in a fluctuating system, though pinning timescales are put in by hand~\cite{erdemci2021effect, dasControlledNeighborExchanges2020a}. Since rosette structures appear frequently during developmental processes~\cite{hardingRolesRegulationMulticellular2014a,trichasMultiCellularRosettesMouse2012a,blankenshipMulticellularRosetteFormation2006}, it is possible that rigidification driven by multi-fold vertex formation is competing with fluctuation-driven fluidization.

\mlm{The effect of persistent fluctuations has also been studied in particle-based glassy systems, with results that are fairly similar to those reported in \ty{CVMs} and \ty{VMs}, except when the fluctuations possess very large persistence times. Interestingly, in that regime the fluctuations become less effective at driving the system from the solid to the fluid phase~\cite{berthierHowActiveForces2017}. It is an open question how these processes -- that can either enhance or inhibit fluidity --  interact with each other to generate tissue dynamics and remodeling.}

One obvious framework that could naturally give rise to both fluctuations and rosette formation is cellular dynamics driven by spatio-temporally fluctuating tension along cell-cell interfaces.  Such fluctuations are regularly observed in experiments~\cite{curranMyosinIIControls2017} and controlled by expression and localization of cytoskeletal and adhesion molecules. For example, fluctuating tension was previously reported for the dynamics of {\it Drosophila} pupal notum, which is a 2D confluent epithelial tissue~\cite{curranMyosinIIControls2017}.  In ref.~28, the authors showed that a 2D CVM with fluctuating tension with some amplitude and persistence time is consistent with the experimental observations. 
\ty{Furthermore, an experimental study combined with a 2D CVM simulation reported that fluctuating tension fluidifies tissues by the intercalation of cells\cite{tetleyTissueFluidityPromotes2019}.}
A study concurrent and independent of the work we report here, by Kranjc~\cite{krajncSolidFluidTransition2020}, analyzed the phase space of fluid-solid transitions in similar CVM models with fluctuating tension. However, it appears that the parameter range of tensions and persistence times studied in that work focuses on the regime where fluidization always dominates over rosette formation. Given experimental observations, this may not be the full experimentally relevant range.  In this work, we extend those previous ideas to characterize how fluctuating tensions across a broad parameter range affect the global tissue mechanics and local cell motion in 2D confluent tissues. We find strongly non-monotonic mechanical response and cell diffusion as a function of the magnitude of the stress fluctuations and their persistence time, consistent with the picture that fluidization due to fluctuations competes with rigidification due to rosette formation.

\section{Results}
We model the dynamics of a 2D confluent tissue using the well-studied 2D \blue{CVM}, where the cells are represented by polygons, and cellular deformations and motions are described by displacements of the vertices and changes in the network topology~\cite{hondaComputerSimulationGeometrical1984}.  
In the 2D CVM, the cellular mechanics and dynamics are governed by the mechanical energy. 
The non-dimensionalized mechanical energy $\epsilon$ of the epithelial tissue is written as a functional of the vertex coordinates $\{\vec{r}_i\}$; 
\begin{eqnarray}
\epsilon(\{\vec{r}_i\}) &=& \cfrac{1}{2} \sum_{\alpha=1}^{N}\left\{k_{\alpha} (a_{\alpha}-a_{0,\alpha})^2
+ (p_{\alpha}-p_{0,\alpha})^2\right\}\nonumber\\&+& \sum_{(i,j)} \Delta{\lambda}_{ij}(t) {\ell}_{ij},
\label{vertex_eq1}
\end{eqnarray}
where $\alpha$ and $N$ denote the label of each cell and the total number of the cells, 
$a_{\alpha}$ and $p_{\alpha}$ are the area and perimeter of cell $\alpha$, and
$a_{0,\alpha}$ and $p_{0,\alpha}$ are the preferred area and perimeter, respectively.
We choose the length scale to satisfy the average cell area $\langle a_{\alpha}\rangle=1$.
$k_{\alpha}$ is the relative area stiffness with respect to the perimeter stiffness of the cell.
Furthermore, we introduce the time-dependent fluctuating part of the tension $\Delta{\lambda}_{ij}(t)$ as the last term in Eq.~(\ref{vertex_eq1}), where ${\ell}_{ij}$ is the edge length between the $i$th and $j$th vertices and the summation runs over the pairs $(i,j)$ of the vertices composing the edges. Based on this mechanical energy, the dynamics of the vertices is described by the following time-evolution equation; 
\begin{eqnarray}
\eta\cfrac{d\vec{r}_i}{dt}&=&-\cfrac{\partial \epsilon(\{\vec{r}_i\})}{\partial \vec{r}_i},\label{timeevolution}
\end{eqnarray}
where $\eta$ is the friction coefficient.

\begin{figure}[t]
 \begin{center}
  \includegraphics[width=85mm]{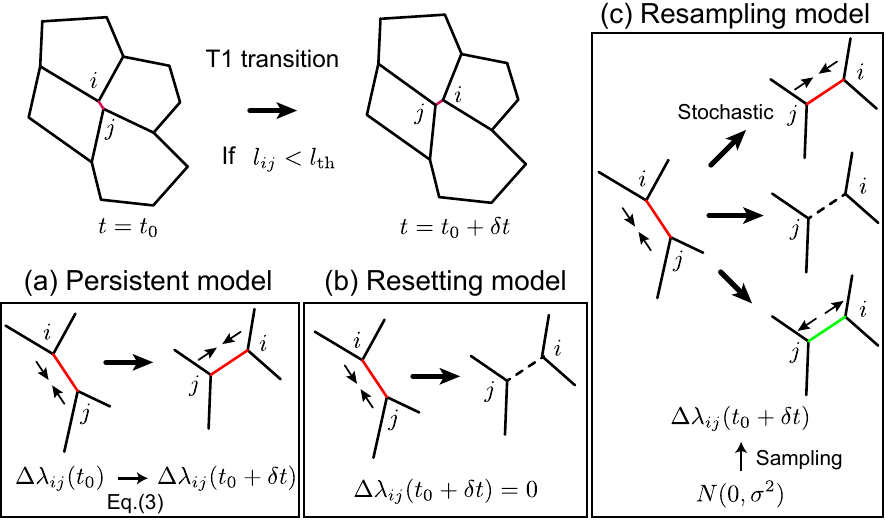}
 \end{center}
 \caption{Schematic illustration of three vertex models with different ways of updating $\Delta\lambda_{ij}$ upon a T1 transition: (a) the persistent model, (b) the resetting model and (c) the resampling model.  In the box for each model, the red solid, the black dashed and the green solid edges show the edges with positive, zero-valued and negative $\Delta\lambda_{ij}$, respectively. Only edges with positive $\Delta\lambda_{ij}$ before a T1 transition are shown, because such edges should tend to shrink more frequently than those with negative $\Delta\lambda_{ij}$. }
 \label{schematic_model}
\end{figure}

We introduce the dynamics of fluctuating part of tension $\Delta{\lambda}_{i j}(t)$ as a general form using a colored Gaussian noise by an Ornstein-Uhlenbeck process~\cite{curranMyosinIIControls2017,krajncSolidFluidTransition2020}: 
\begin{eqnarray}
\cfrac{d \Delta\lambda_{i j}(t)}{dt}=-\cfrac{\Delta\lambda_{i j}(t)}{\tau}+\xi_{ij}(t),\label{linetension}
\end{eqnarray}
where $\xi_{ij}(t)$ is a white Gaussian noise satisfying  $\langle\xi_{ij}(t)\rangle=0$ and
$\langle {\xi}_{i j}(t_1) {\xi}_{kl}(t_2) \rangle = 2\sigma^2/\tau\delta_{ik}\delta_{jl}\delta(t_1-t_2)$. 
Here, $\Delta{\lambda}_{i j}(t)$ satisfies $\langle\Delta\lambda_{ij}(t)\rangle=0$ and
$\langle \Delta{\lambda}_{i j}(t_1) \Delta{\lambda}_{kl}(t_2) \rangle = \delta_{ik}\delta_{jl}\sigma^2 e^{-|t_1-t_2|/{\tau}}$. 
The characteristic time scale of the fluctuating tension is determined by the persistence time $\tau$, and $\sigma$ sets the characteristic amplitude.

\begin{figure*}[hbt]
  \includegraphics[width=178mm]{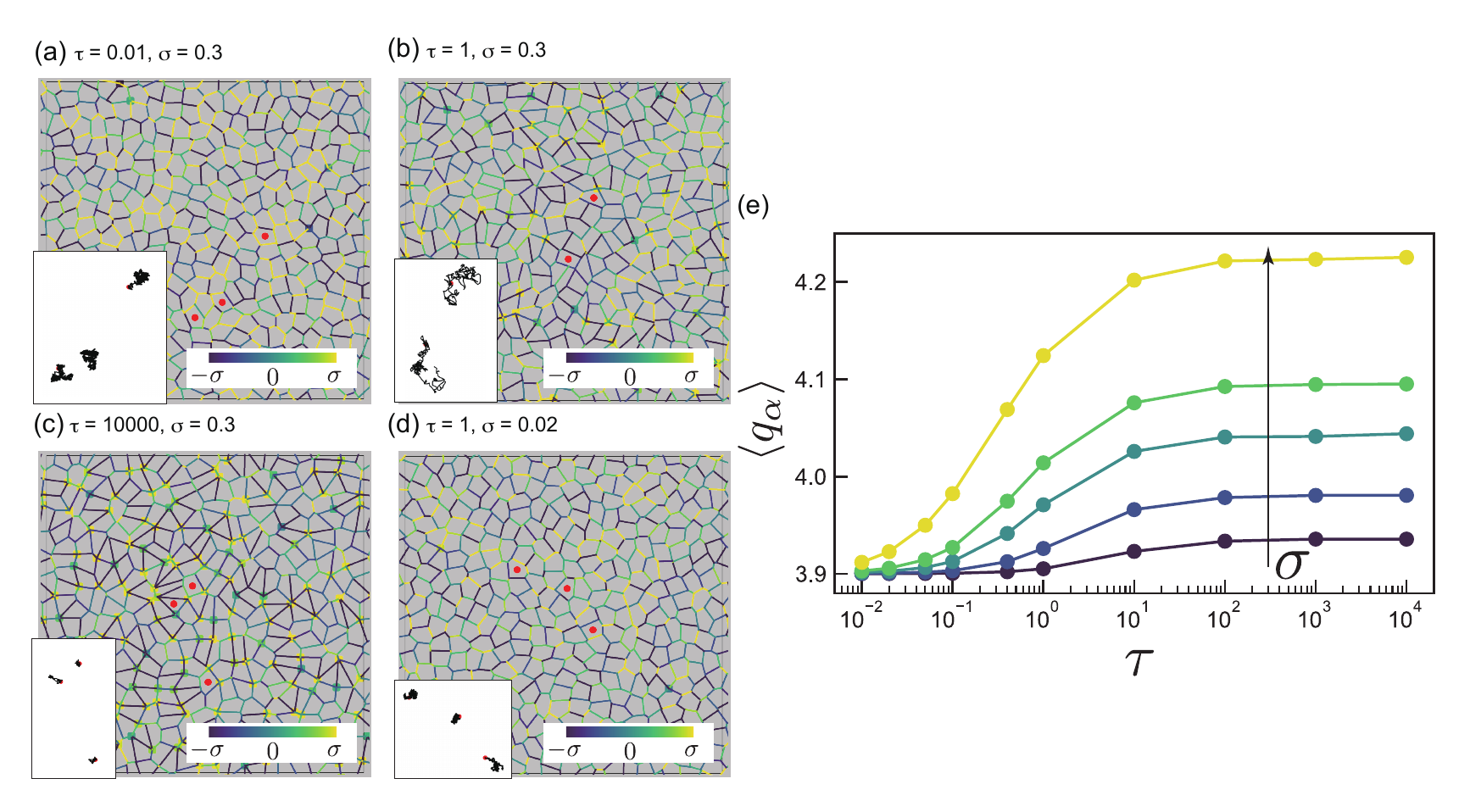}
 \caption{Snapshots of the cellular configurations obtained in our numerical simulation with different $\tau$ and $\sigma$: (a) $\tau=0.01, \sigma = 0.3$, (b) $\tau=1, \sigma = 0.3$, (c) $\tau=10000, \sigma = 0.3$, (d) $\tau=1, \sigma = 0.02$. We draw the edges with the color mapping the value of $\Delta\lambda_{ij}$. The color is mapped to $\Delta\lambda_{ij}$ ranging in $[-\sigma,\sigma]$ using the color map in each figure. If $\Delta\lambda_{ij} < -\sigma \ (\Delta\lambda_{ij}>\sigma)$, we color the edge with the color corresponding to $\Delta\lambda_{ij} =-\sigma$ ($\Delta\lambda_{ij} = \sigma$). 
 The trajectories of cells marked with red circles are also shown with black solid lines in the insets. We show the trajectories in the interval of $3000$ time unit for (a)(c)(d) and $300$ time unit for (b).
 In (a-d), we also highlight the trapped edges defined in the main text with the squares colored using the same color map as the other edges.
(e) Average cell shape index $\langle q_{\alpha}\rangle$ for $p_0=3.90$. $\sigma \in [ 0.02, 0.05, 0.10,0.15, 0.30]$ (from dark color to light color).
 }
 \label{snapshot}
\end{figure*}

In this study, we investigate 
the effect of 
$\sigma$ and 
$\tau$ 
on the cellular dynamics. 
In our numerical simulation, we solve Eq.~(\ref{timeevolution}) using \blue{the} forward Euler method with a time step $\delta t=0.01$. 
We set $k_{\alpha}=1$ and $\eta=1$.
We initially prepare \blue{a hexagonal pattern of} 340 cells in a squared area with periodic boundary conditions, then run the simulation with a large amplitude of fluctuation in tension to randomize the cellular configuration for $100\ \rm{natural\ time\ units}$ ($\sigma=0.35,\tau=1$). 
After the randomization, we simulate dynamics in the system with the target values of $\sigma$ and $\tau$ for $10^4$ natural time units to initialize the system, then report dynamical data over an additional $10^5$ natural time units.  
We perform T1 transitions by flipping edges with a length below a threshold $l_{\rm th}$ in clockwise direction by $90^{\circ}$, if the energy decreases after the T1 transition. 
We set $l_{\rm th}$ to $5\%$ of the length $l_{\rm hex}=\sqrt{2\sqrt{3}}/3$ of an edge of a regular hexagonal cell with area $1$. 

Unfortunately, there is little experimental data describing how tensions evolve after a T1 transition.  In the absence of such data, one could envision several scenarios for how to specify the tension on the newly formed edge.  We consider three options in this manuscript, illustrated schematically in Fig.~\ref{schematic_model}.  In the first  “persistent model", we keep the same value of the tension  $\Delta{\lambda}_{i j}(t)$ after the T1 transition as was on the shrinking edge before the T1 transition. In the second  “resetting model", $\Delta{\lambda}_{i j}(t)$ along the new edge is set to zero after the T1 transition.  In the last “resampling model", we resample $\Delta{\lambda}_{i j}(t)$ from the normal distribution $N(0,\sigma^2)$ with zero mean and variance $\sigma^2$, which is the stationary distribution of the Ornstein-Uhlenbeck process described by Eq.~(\ref{linetension}). 

\ty{Our first set of results focus on the persistent model, as there are some minimal observations in the literature that are consistent with it. For example, Bosveld et al. reported that increasing tension at cell edges causes the accumulation of F-actin binding protein Vinculin at the tri-cellular junctions (TCJs), while reducing tension decreases the amount of active Myosin II at the TCJs\cite{bosveldTricellularJunctionsHot2018}. Furthermore, Tricellulin, which is a protein localizing at TCJs, recruits the Cdc42 GEF Tuba, which activates Cdc42 to promote the assembly of an actomyosin meshwork at the TCJs as well as bicellular junctions. This suggests that there could be a positive correlation between the activity of TCJs and the edge tension, and that TCJs may retain memory of the edge tension before the T1 transition. Also, there is some experimental evidence for a correlation in myosin intensity before and after a T1 transition\cite{curranMyosinIIControls2017}. We discuss the resetting and resampling models later in the manuscript.}

First, we study the qualitative effect of varying the overall magnitude $\sigma$ and persistence $\tau$ of stress fluctuations on cellular structure. Snapshots of the cellular configuration from the numerical simulations for different sets of $\sigma$ and $\tau$ for fixed $p_0=3.9$ are shown in Fig.~\ref{snapshot}.  For fixed $\sigma=0.3$ (Fig.~\ref{snapshot}(a-c)), we found that the cellular shape is more irregular for larger $\tau$, while larger $\sigma$ gives more irregular cell shapes for fixed $\tau=1$ (Fig.~\ref{snapshot}(b) and (d)).

Figure~\ref{snapshot}(e) quantifies the cell shape index $q_{\alpha}=p_{\alpha}/\sqrt{a_{\alpha}}$, which tends to increase when the cellular shape is anisotropic or the number of edges composing the cell is large. This panel confirms that cell shape index increase with increasing $\tau$ and $\sigma$. This is not surprising, as increasing  $\tau$ and $\sigma$ increases the number of persistently shrinking (large positive $\Delta\lambda_{ij}$) and expanding (large negative $\Delta\lambda_{ij}$) edges. 

There is one surprise. Although previous work in vertex models has identified a strong correlation between cell shape and tissue fluidity, it is clear from the insets of Fig.~\ref{snapshot}(a-c) illustrating cell trajectories that there is a non-monotonic behavior for cell diffusivity as a function of increasing $\tau$, despite the fact that cell shapes become more irregular with increasing $\tau$.  Similarly, Fig.~\ref{snapshot}(e) illustrates that there is a small-$\tau$ regime where the cell shape depends sensitively on $\tau$, and a large $\tau$ regime where cell shape becomes almost independent of $\tau$. Moreover, at these larger values of $\tau$, irregular cell shapes coexist with many very short edges, highlighted with square symbols in Fig.~\ref{snapshot}, a point we will return to later. 

\begin{figure}
 \begin{center}
  \includegraphics[width=80mm]{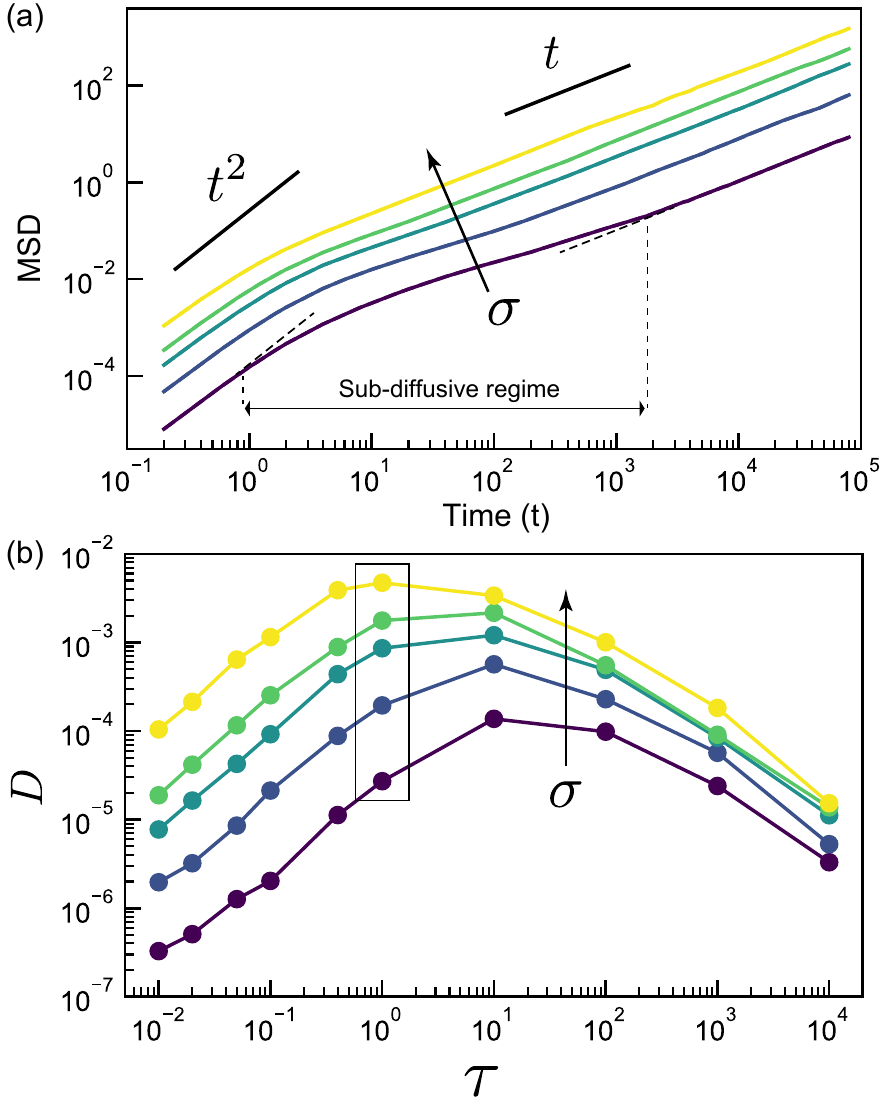}
 \end{center}
 \caption{Cell dynamics as a function of the magnitude of stress fluctuations $\sigma$ and the persistence of stress fluctuations $\tau$. (a) Mean-squared displacement (MSD) as a function of time for different $\sigma$ with fixed $\tau=1$ are shown. The solid lines are guides for eyes indicating the power laws $t$ and $t^2$, respectively. \blue{The subdiffusive regime is also indicated for $\sigma=0.02$.} 
(b) Long-time diffusion constant $D$ extracted from the MSD as a function of $\tau$ for different $\sigma$. \blue{The box highlights $D$ values corresponding to $\tau =1$, derived from the data shown in panel (a).} In (a) and (b), $\sigma \in [ 0.02, 0.05, 0.10,0.15, 0.30]$ (from dark color to light color).}
 \label{msd}
\end{figure}

To quantitatively characterize the cellular dynamics and begin to understand the origin of the observed non-monotonic behavior, we calculate the mean-squared displacement (MSD) of the area centroid of the cells.  Example cell trajectories are shown in the insets to Fig.~\ref{snapshot}. 
In Fig.~\ref{msd}(a), we show the MSD curves as a function of the time $t$ for $\tau=1$ and $p_0=3.9$. \mlm{Previous work has demonstrated that at zero temperature this model transitions to a fluid-like state for $p_0 > p_0^* \sim 3.8$, so that this system is in a fluid-like phase, albeit with glassy dynamics. Different values of $p_0$ are explored below.}

The curves exhibit ballistic behavior with $\rm{MSD}\sim t^2$ at short time scales $t\ll\tau$. 
At long time scales \blue{$t \gg \tau$}, the cellular dynamics exhibit diffusive behavior with $\rm{MSD}\sim t$. 
Notably, we found that the MSD exhibits a sub-diffusive \blue{regime} characterized by $\rm{MSD}\sim t^{\alpha}$ with $0<\alpha<1$ at intermediate time scales. This \blue{subdiffusive regime}, also seen in CVM simulations with Brownian noise on the vertices~\cite{sussmanAnomalousGlassyDynamics2018a}, is a characteristic feature of glasses and indicates that cells are being caged by \blue{their} neighbors at intermediate timescales. The \blue{subdiffusive regime} becomes less prominent at large values of $\sigma$, suggesting that $\sigma$ is playing a role similar to an effective temperature, where the system becomes more fluid-like as $\sigma$ increases the overall level of fluctuations. 

We further characterize the dynamics by estimating the diffusion constant $D=\lim_{t\to \infty}{{\rm MSD}(t)/4t}$ for values of $t>10^{4}$ for different $\tau$ and $\sigma$ as shown in Fig.~\ref{msd}(b). 
The diffusion constant $D$ exhibits non-monotonic dependence on $\tau$, where $D$ is maximized at intermediate $\tau \sim 1 - 10$.  This quantitatively confirms the cellular dynamics exhibits two different regimes at small and large $\tau$, respectively, which we discuss in detail below. 

\begin{figure}
 \begin{center}
  \includegraphics[width=80mm]{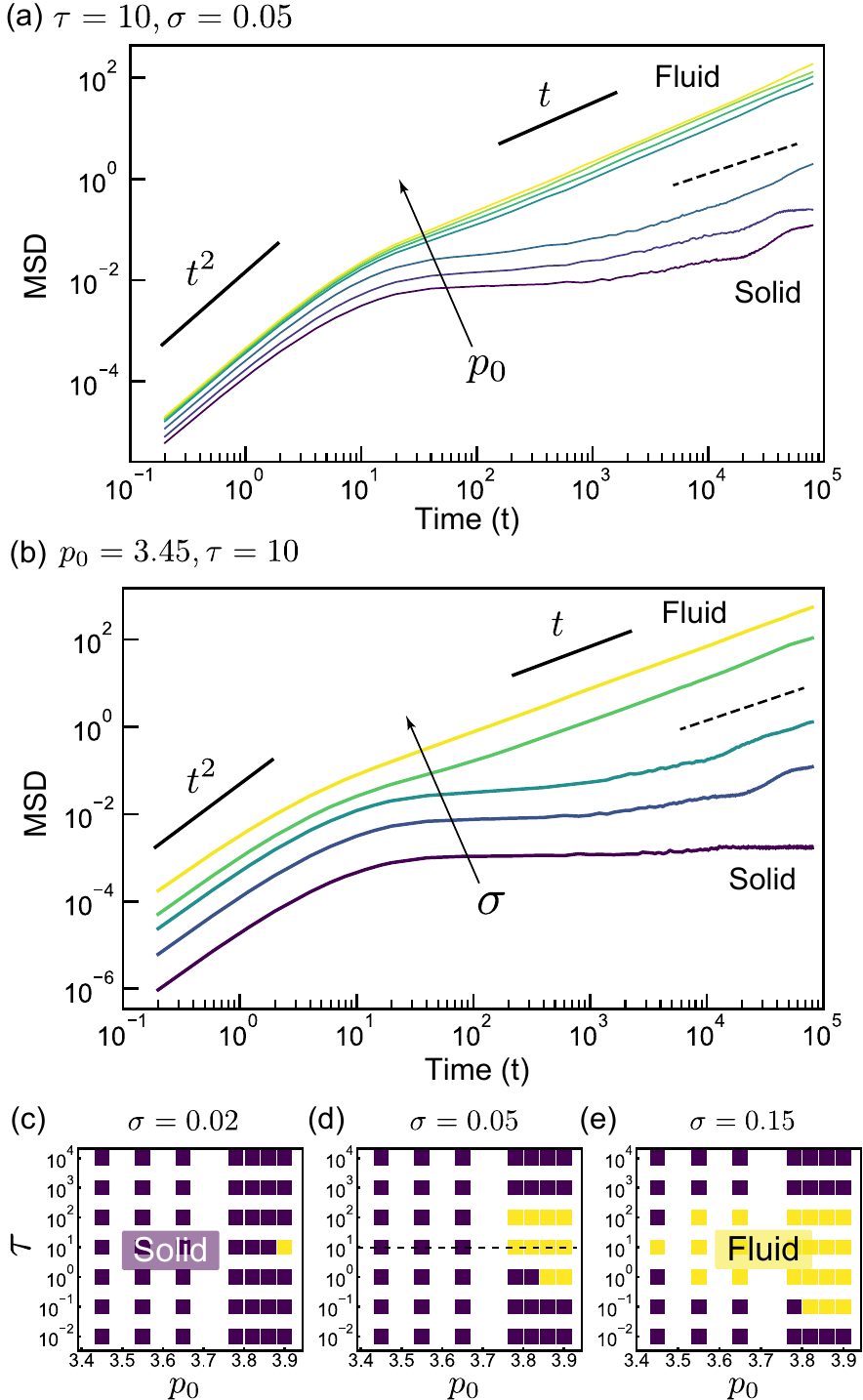}
 \end{center}
 \caption{Glassy behaviors of the confluent tissues for different $p_0$, $\tau$ and $\sigma$. (a)(b) The MSD curves for fixed $(\tau=10,\sigma=0.05)$ and $(p_0=3.45,\tau=10)$ are shown. The dashed lines distinguish between the fluid-like and the solid-like phases. The solid lines are guides for eyes indicating the power law $t$ and $t^2$, respectively. In (a), $p_0 \in [3.45, 3.55,3.65,3.78,3.82,3.86,3.90]$ (from dark color to light color). In (b), $\sigma \in [ 0.02, 0.05, 0.10,0.15, 0.30]$ (from dark color to light color). (c-e) Phase diagrams for the glassy behaviors. The solid-like ($D<D^*$) and fluid-like ($D>D^*$) tissues are shown as purple and yellow data points, respectively. The dashed line in (d) is the scanning line for the data in (a).}
 \label{PhaseDiagram}
\end{figure}

We also investigate the effect of shape index $p_0$, another parameter which is known to control the rigidity of the tissue~\cite{parkUnjammingCellShape2015,biDensityindependentRigidityTransition2015a,biMotilityDrivenGlassJamming2016a}. 
Previous 2D CVM studies showed that the confluent tissue becomes solid-like (fluid-like) for small (large) $p_0$ with the transition point $p_0^*\sim3.81$~\cite{parkUnjammingCellShape2015,biMotilityDrivenGlassJamming2016a}.
In Fig.~\ref{PhaseDiagram}(a), we show the MSD curves for different $p_0$ with fixed $\tau=10$ and $\sigma=0.05$. 
For small $p_0$, the MSD curves show ballistic behaviors at short time scales, plateaus at intermediate time scales and diffusive behaviors at long time scales, indicating fluidity at the longest timescales.  Again, the plateau indicating glassy dynamics is less prominent for large $p_0$, confirming that the tissue becomes less glassy and more fluid-like as $p_0$ increases. 
Figure~\ref{PhaseDiagram} also shows the MSD curves for different $\sigma$ with fixed $p_0=3.45$ and $\tau=10$. 
Since $p_0=3.45$ is well below $p_0^*$, for small $\sigma$ the tissue is solid-like, exhibiting non-diffusive behavior at long timescales, but increasing $\sigma$ leads to diffusion at long timescales, indicating fluidization of the tissue.

Therefore, in our model, the trio of parameters $[ p_0, \sigma, \tau ]$ controls the fluid-to-solid transition. \mlm{While the shear modulus is a natural metric for the fluid-to-solid transition in systems without fluctuations~\cite{merkelMinimallengthApproachUnifies2019},  subtleties arise in thermalized or active systems because calculating the shear modulus requires taking the limit of infinitely slow driving. In the glassy physics community, therefore, a solid is usually defined as a system where the viscosity is larger than an arbitrary threshold. Previous work~\cite{biMotilityDrivenGlassJamming2016a,czajkowskiGlassyDynamicsModels2019a} has demonstrated that a similar metric, namely a threshold on the diffusivity, accurately distinguishes between solid-like systems where the cells largely remain within a cage of their neighbors and fluid-like systems where cells regularly exchange neighbors. 
The dashed lines in Fig.~\ref{PhaseDiagram}(a)-(b) correspond to a threshold in measured diffusivity of $D^* = 10^{-4}$,  illustrating that, for the system parameters we study in this work, this choice of threshold does indeed distinguish between systems with a significant sub-diffusive plateau (e.g. cells trapped by a cage of neighbors) and those with no such plateau (e.g. cells changing neighbors). 
Therefore, we define the fluid-solid transition by a threshold in the magnitude of the diffusion constant $D^* = 10^{-4}$.} In Fig.~\ref{PhaseDiagram}(c-e), we show the cross-sections of the three dimensional (3D) phase diagram of solid-to-fluid transition with respect to these parameters. As highlighted in Fig.~\ref{PhaseDiagram} (d) and (e), 
there is always a re-entrant fluid-solid transition as a function of $\tau$.

\begin{figure}[t]
 \begin{center}
  \includegraphics[width=85mm]{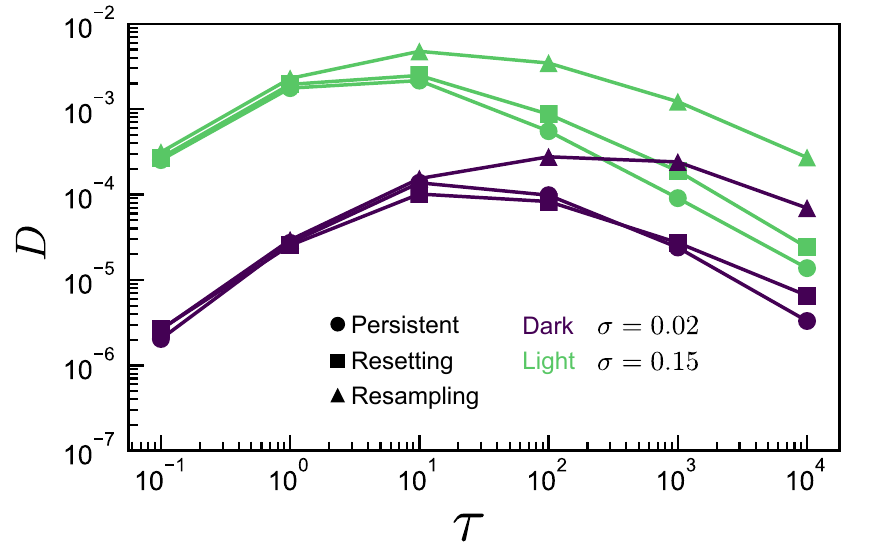}
 \end{center}
 \caption{Comparison of $D$ between three models: the persistent model (circle), the resetting model (square), the resampling model (triangle). $D$ vs. $\tau$ for the three models, respectively. Dark and light markers represent the data with $\sigma=0.02$ and $\sigma=0.15$, respectively.
The distributions $F(l^i)$ were too broad to determine the threshold $l^*$ for the trapping edges in the following data points: $(\tau,\sigma)=(0.1,0.02)$ in the persistent model, $(\tau,\sigma)=(0.1,0.02/0.15),(1,0.02)$ in the resetting model, $(\tau,\sigma)=(0.1,0.02/0.15),$ $(1,0.02/0.15),$ $(10,0.02/0.15),$ $(100,0.02),$ $(1000,0.02),$ $(10000,0.02)$ in the resampling model.
We hence set $l^*=l_{\rm th}$ for these data points.
 }
 \label{3model_result}
\end{figure}

\mlm{One obvious question is whether our results depend strongly on our choice of how to resample the tension in the newly created edges after a T1 swap.  The “persistent'' model we have considered so far gives the new edge after a T1 swap the same tension as the old edge, which will clearly favor trapped edges where the tension is larger and contractile, since edges with large tension are likely to keep shrinking. Therefore, we also investigate more “democratic'' ways of sampling tensions in the new T1 edge, illustrated schematically in Fig 1 (b) and (c), which we term “resetting'' and “resampling'' models. Figure~\ref{3model_result} shows that the resetting and resampling models generate the same diffusion constants as the persistent models in the small-$\tau$ regimes, consistent with the hypothesis that fluctuation-driven diffusion, which should be the same in all models, dominates at low $\tau$. In addition, there is still non-monotonic behavior in all three models, with the diffusion constant decreasing at large $\tau$. }
 
To investigate the mechanisms driving this re-entrant behavior, we focus on the \blue{persistent model} with $p_0=3.9$, since the re-entrant behavior is observed for many $p_0$ values. We first focus on the small-$\tau$ regime, where increasing $\tau$ increases diffusion. Since cell diffusion is driven by cell rearrangements that occur when a T1 edge shrinks to zero, we first \blue{consider} the characteristic timescale required for an edge of length $l$ to shrink to zero. This is not a straightforward first-passage-time problem, however, as the edges in the tessellation cannot grow towards positive infinity. When the length rises significantly above unity, a T1 transition in a neighboring edge is likely to be triggered, generating a complicated absorbing boundary condition.  

\blue{Ultimately}, we are interested in the diffusion of a cell's center of mass. We anticipate that when an edge shrinks to zero and experiences a T1 transition, the cell center displaces a characteristic fraction of the distance over which the edge shrinks. 
The cell center-of-mass displacements can be approximated as a memory-less chain of edge-shrinking events.
Therefore, rather than focusing on \blue{first-passage} time statistics, we study the mean-field behavior of an edge length, and calculate the characteristic timescale over which it diffuses in the absence of any boundary conditions. In section I in ESI, we also perform a numerical study of a related first-passage-time problem and demonstrate that it also exhibits the same scaling in the small-$\tau$ regime as described below.

\begin{figure}[htb]
 \begin{center}
  \includegraphics[width=85mm]{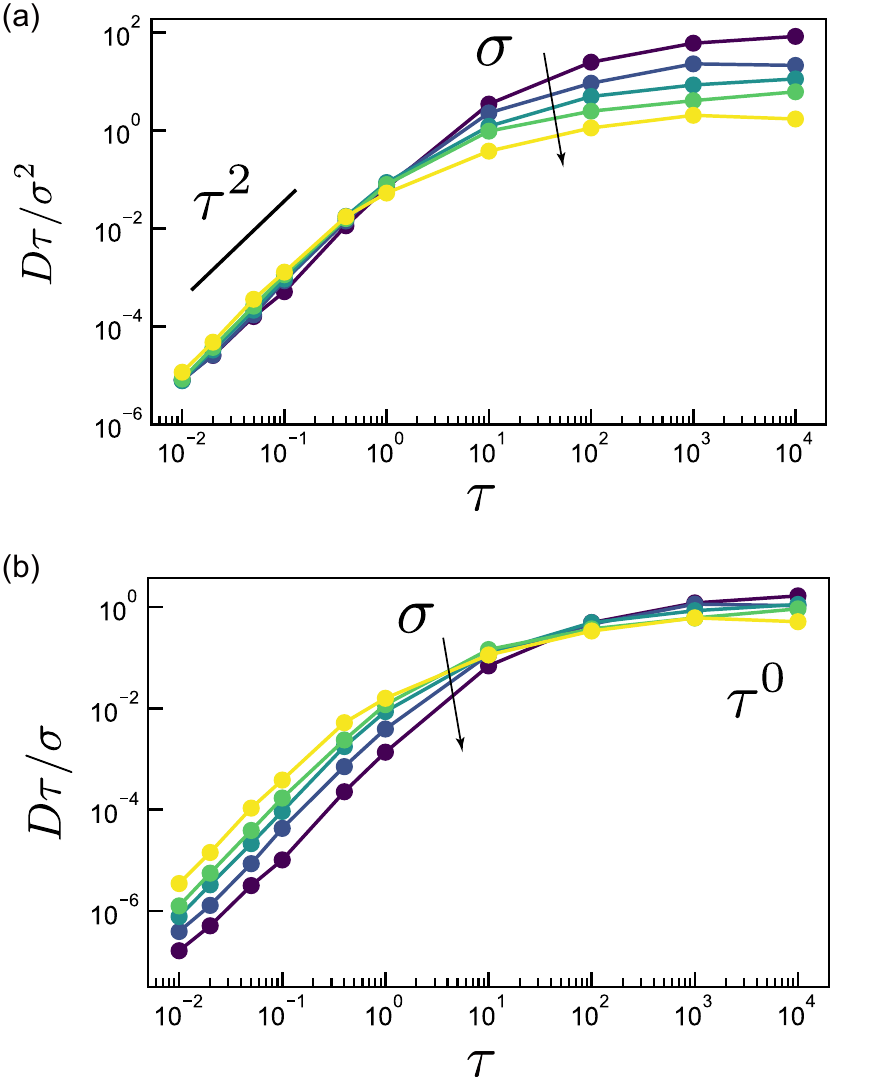}
 \end{center}
 \caption{Data collapse in a plot of $D\tau/\sigma^2$ vs. $\tau$ demonstrates the scaling relation $D\propto\sigma^2\tau$ in the small $\tau$ regime. The solid line is a guide to the eye indicating the power law $\tau^2$. $\sigma \in [ 0.02, 0.05, 0.10,0.15, 0.30]$ (from dark color to light color).
Data collapse in a plot of $D\tau/\sigma$ vs. $\tau$ demonstrates the scaling relation $D\propto\sigma/\tau$ in the large $\tau$ regime.
 $\sigma \in [ 0.02, 0.05, 0.10,0.15, 0.30]$ (from dark color to light color).}
 \label{scaling}
\end{figure}

Assuming that the tension  of the edge is determined only by the fluctuating part of the tension $\Delta\lambda$,
we obtain the following time-evolution equation for the edge length $l$; 
\begin{eqnarray}
\frac{d l}{dt} = -\Delta\lambda,\label{2VM}
\end{eqnarray}
where the time-evolution of $\Delta\lambda$ is given by Eq.~(\ref{linetension}) with $l_{ij}=l$ and $\Delta\lambda_{ij}=\Delta\lambda$. 
Then the time evolution of the MSD of $l$ is: ${\rm MSD}_{l}(t) = 2\sigma^2\tau t + 2\sigma^2\tau^2(\exp(-t/\tau)-1)$~\cite{bodrovaUnderdampedScaledBrownian2016}. 
Accordingly, when $t \gg \tau$, the MSD of $l$ scales as ${\rm MSD}_l(t) = 2 D_l t$, where $D_{l}=\sigma^2\tau$ is the diffusion constant of the edge length $l$.
If we assume that this is the primary timescale driving cell rearrangements as discussed above, then we predict the total diffusion rate is simply $D\sim D_l =\sigma^2\tau$.  This is in good agreement with numerical data for the small-$\tau$ regime as shown in Fig.~\ref{scaling}. This confirms that in this regime, the fluidization generated by increasing $\tau$ occurs because edges shrink more persistently.

This argument obviously breaks down in the large-$\tau$ regimes ($\tau >\sim 1$), where we observe that the diffusion constant decreases with increasing $\tau$.  

\mlm{One way the argument could break down is that cells no longer rearrange when edges shrink to zero length, resulting in “trapped" edges, or rosette structures, where more than three cells meet. A CVM study by Yan {\it et al.} in the limit of zero fluctuations recently showed that rosette structures can rigidify the epithelial tissue~\cite{yanMulticellularRosettesDrive2019}.  Although our model strictly only contains 3-fold coordinated vertices, in the persistent model we indeed observe a large number of very short edges at large $\tau$ as shown in Fig.~\ref{snapshot}(a-d) and Fig.~S5(b) in ESI, and it could be that such short edges are constraining the structure in a  manner similar to rosettes.  }

\mlm{Alternatively, our argument could also break down if the rate-limiting step is {\bf not} the time it takes an edge to shrink in the presence of unbalanced forces. This could happen, for example, if the tension network evolves so slowly that it takes a long time to achieve a state with unbalanced forces.}

\mlm{To test the first possibility, we performed a thorough analysis of both the number of trapped edges and the time over which such edges remained trapped, discussed in detail in the ESI.  While we find a significant increase in trapped edges in the persistent model (Fig.~\ref{snapshot}(a-d)), we find only a small increase in the resetting model and almost no change in the resampling model for small force amplitudes.  This indicates that trapping of edges is not the dominant mechanism contributing to the non-monotonic behavior of the diffusivity (Figs.~S7 and S8 in ESI).}

\textcolor{black}{Fig.~\ref{scaling}(b) shows a scaling collapse of the large $\tau$ regime of the persistent model.  This scaling demonstrates that $D$ asymptotically approaches $D\sim\sigma/\tau$, and Fig.~S3 in ESI demonstrates similar scaling at large $\tau$ in the resetting and resampling models.   As the diffusion constant is the rate at which the system rearranges to explore new configurations, this indicates that new configurations are being explored at precisely the same rate that the tension network is being remodeled, independent of the details of T1 rearrangements.  This in turn suggests that there may be a separation of timescales, so that a rearrangement quickly allows the system to find a slow, nearly force-balanced state, which becomes unbalanced again over a timescale $\tau$ and generates a new rearrangement.}  

\textcolor{black}{Unfortunately, even in non-active disordered glasses, identifying such a separation of timescales directly in simulations is notoriously difficult due to the presence of avalanches and long-range elastic interactions~\cite{nicolasDeformationFlowAmorphous2018}. Nevertheless, the fact that a similar scaling with persistence time $\tau$ is seen in a very different glassy simulation of active Ornstein-Uhlenbeck particles~\cite{berthierHowActiveForces2017, debetsCageLengthControls2021} suggests that $\tau$ generically sets the timescale for the diffusion dynamics when it is larger than any other relaxation timescale in the problem.}

 \section{Discussion}
 Taken together, these results suggest that in tissues with fluctuating tensions, \mlm{there is a fast-fluctuation regime dominated by the time it takes an edge to shrink, and a slow-fluctuation regime dominated by the slow evolution of the tension network.}  In general, increasing the magnitude of the tension always increases the fluidity of the tissue, while increasing the persistence of fluctuations has a non-monotonic impact on tissue fluidity.  For short persistence times, the diffusivity is dominated by fluctuations and increases with increasing persistence. We confirm this by predicting and demonstrating a scaling collapse of our data in this regime. \ty{In contrast, for larger persistence times the cell dynamics are governed by the persistence time itself, suggesting they are slaved to the slow dynamics of the tension in the network.}
 
 Our results in the small-$\tau$ regime are entirely consistent with independent work recently published by Kranjc~\cite{krajncSolidFluidTransition2020}, which 
 found simple monotonic relationships between $\sigma$, $\tau$, and the diffusivity in this regime.  
 
 However, it is reasonable to expect that fluctuations in stress, generated by correlated and cooperative localization of large number of cytoskeletal molecules, may persist longer than the natural time unit in these simulations, which roughly corresponds to the time required for cells to find a new stable state after executing a T1 transition. For example, rough estimates for rearrangement timescales from experiments in {\it Drosophila} are typically less than 10 minutes~\cite{kaszaSpatiotemporalControlEpithelial2014a}, while fluctuations in tensions due to mechanisms like planar cell polarity can last upwards of 30 minutes~\cite{kaszaSpatiotemporalControlEpithelial2014a}, and multi-fold coordinated vertices are often observed in such systems.  Therefore, the large-$\tau$ regime, \blue{explored in vertex models} for the first time, is likely to be relevant for many experiments.

 \mlm{One open question is whether the trapped edges that we observed in the persistent model are contributing to the rigidification for that model.} Strictly speaking, the constraint counting argument developed in ref.~23 depends on the fact that multi-fold coordinated vertices explicitly reduce the number of degrees of freedom available to the system. This is not the case for our effective multi-fold coordinated vertices, where the total number of degrees of freedom remains constant.  On the other hand, very short, high tension edges do place strong constraints on the dynamics of the attached vertices. As shown by some of us in ref.\blue{~41}, such short edges in systems with heterogeneous tensions can generate cusps in the potential energy landscape that can trap vertices\blue{~\cite{sussmanSoftSharpInterfaces2018b}}.  Therefore, future work could focus on using some of these ideas to generalize the static arguments made in ref.~23 to explain enhanced rigidity in dynamic systems. In particular, it would be interesting to know what sets the characteristic lengthscale for trapped edges, and whether it depends on an effective temperature driving fluctuations.
 
 From an experimental perspective, our work clarifies that fluctuating tensions can drive either fluidization or rigidity depending on the parameter regime. Given that the tension dynamics just after T1 transitions play an important role in this balancing act, it would be especially useful to gather data, using tools such as laser ablation or optogenetics, about how these tensions evolve in different {\it in vivo} and {\it in vitro} systems.  As the rigidity/fluidity of biological tissues can help set timescales for processes like body axis elongation~\cite{wangAnisotropyLinksCell2020, mongeraFluidtosolidJammingTransition2018} or wound healing, it could be that organisms tune the magnitude or persistence time of stress fluctuations to control such processes. It would be interesting to look for such trends in model organisms.

\section*{Conflicts of interest}
There are no conflicts to declare.

\begin{acknowledgments}
T.Y. thank Kyogo Kawaguchi, Kyosuke Adachi and Yosuke Fukai for fruitful discussion.
This work is supported by Grant-in-Aid for JSPS Fellows (Grant No. 18J01239), KAKENHI Grant No. 17H07366 and RIKEN HOKUSAI supercomputer systems (Project Q19433) to T.Y. 
M.L.M. and D.M.S. acknowledge support from Simons Foundation Grants \#46222 and \#454947 and NSF-PHY-1607416.
\end{acknowledgments}

\bibliography{rsc}

\begin{thebibliography}{41}%
\makeatletter
\providecommand \@ifxundefined [1]{%
 \@ifx{#1\undefined}
}%
\providecommand \@ifnum [1]{%
 \ifnum #1\expandafter \@firstoftwo
 \else \expandafter \@secondoftwo
 \fi
}%
\providecommand \@ifx [1]{%
 \ifx #1\expandafter \@firstoftwo
 \else \expandafter \@secondoftwo
 \fi
}%
\providecommand \natexlab [1]{#1}%
\providecommand \enquote  [1]{``#1''}%
\providecommand \bibnamefont  [1]{#1}%
\providecommand \bibfnamefont [1]{#1}%
\providecommand \citenamefont [1]{#1}%
\providecommand \href@noop [0]{\@secondoftwo}%
\providecommand \href [0]{\begingroup \@sanitize@url \@href}%
\providecommand \@href[1]{\@@startlink{#1}\@@href}%
\providecommand \@@href[1]{\endgroup#1\@@endlink}%
\providecommand \@sanitize@url [0]{\catcode `\\12\catcode `\$12\catcode
  `\&12\catcode `\#12\catcode `\^12\catcode `\_12\catcode `\%12\relax}%
\providecommand \@@startlink[1]{}%
\providecommand \@@endlink[0]{}%
\providecommand \url  [0]{\begingroup\@sanitize@url \@url }%
\providecommand \@url [1]{\endgroup\@href {#1}{\urlprefix }}%
\providecommand \urlprefix  [0]{URL }%
\providecommand \Eprint [0]{\href }%
\providecommand \doibase [0]{http://dx.doi.org/}%
\providecommand \selectlanguage [0]{\@gobble}%
\providecommand \bibinfo  [0]{\@secondoftwo}%
\providecommand \bibfield  [0]{\@secondoftwo}%
\providecommand \translation [1]{[#1]}%
\providecommand \BibitemOpen [0]{}%
\providecommand \bibitemStop [0]{}%
\providecommand \bibitemNoStop [0]{.\EOS\space}%
\providecommand \EOS [0]{\spacefactor3000\relax}%
\providecommand \BibitemShut  [1]{\csname bibitem#1\endcsname}%
\let\auto@bib@innerbib\@empty
\bibitem [{\citenamefont {Ariga}\ \emph {et~al.}(2018)\citenamefont {Ariga},
  \citenamefont {Tomishige},\ and\ \citenamefont
  {Mizuno}}]{arigaNonequilibriumEnergeticsMolecular2018}%
  \BibitemOpen
  \bibfield  {author} {\bibinfo {author} {\bibfnamefont {T.}~\bibnamefont
  {Ariga}}, \bibinfo {author} {\bibfnamefont {M.}~\bibnamefont {Tomishige}}, \
  and\ \bibinfo {author} {\bibfnamefont {D.}~\bibnamefont {Mizuno}},\
  }\href@noop {} {\bibfield  {journal} {\bibinfo  {journal} {Physical Review
  Letters}\ }\textbf {\bibinfo {volume} {121}},\ \bibinfo {pages} {218101}
  (\bibinfo {year} {2018})}\BibitemShut {NoStop}%
\bibitem [{\citenamefont {Ariga}\ \emph {et~al.}(2020)\citenamefont {Ariga},
  \citenamefont {Tomishige},\ and\ \citenamefont
  {Mizuno}}]{arigaExperimentalTheoreticalEnergetics2020}%
  \BibitemOpen
  \bibfield  {author} {\bibinfo {author} {\bibfnamefont {T.}~\bibnamefont
  {Ariga}}, \bibinfo {author} {\bibfnamefont {M.}~\bibnamefont {Tomishige}}, \
  and\ \bibinfo {author} {\bibfnamefont {D.}~\bibnamefont {Mizuno}},\
  }\href@noop {} {\bibfield  {journal} {\bibinfo  {journal} {Biophysical
  Reviews}\ }\textbf {\bibinfo {volume} {12}},\ \bibinfo {pages} {503}
  (\bibinfo {year} {2020})}\BibitemShut {NoStop}%
\bibitem [{\citenamefont
  {Hayashi}(2018)}]{hayashiApplicationFluctuationTheorem2018}%
  \BibitemOpen
  \bibfield  {author} {\bibinfo {author} {\bibfnamefont {K.}~\bibnamefont
  {Hayashi}},\ }\href@noop {} {\bibfield  {journal} {\bibinfo  {journal}
  {Biophysical Reviews}\ }\textbf {\bibinfo {volume} {10}},\ \bibinfo {pages}
  {1311} (\bibinfo {year} {2018})}\BibitemShut {NoStop}%
\bibitem [{\citenamefont {Yanagida}\ \emph {et~al.}(2008)\citenamefont
  {Yanagida}, \citenamefont {Iwaki},\ and\ \citenamefont
  {Ishii}}]{yanagidaSingleMoleculeMeasurements2008}%
  \BibitemOpen
  \bibfield  {author} {\bibinfo {author} {\bibfnamefont {T.}~\bibnamefont
  {Yanagida}}, \bibinfo {author} {\bibfnamefont {M.}~\bibnamefont {Iwaki}}, \
  and\ \bibinfo {author} {\bibfnamefont {Y.}~\bibnamefont {Ishii}},\
  }\href@noop {} {\bibfield  {journal} {\bibinfo  {journal} {Philosophical
  Transactions of the Royal Society B: Biological Sciences}\ }\textbf {\bibinfo
  {volume} {363}},\ \bibinfo {pages} {2123} (\bibinfo {year}
  {2008})}\BibitemShut {NoStop}%
\bibitem [{\citenamefont {Toyabe}\ \emph {et~al.}(2011)\citenamefont {Toyabe},
  \citenamefont {{Watanabe-Nakayama}}, \citenamefont {Okamoto}, \citenamefont
  {Kudo},\ and\ \citenamefont
  {Muneyuki}}]{toyabeThermodynamicEfficiencyMechanochemical2011}%
  \BibitemOpen
  \bibfield  {author} {\bibinfo {author} {\bibfnamefont {S.}~\bibnamefont
  {Toyabe}}, \bibinfo {author} {\bibfnamefont {T.}~\bibnamefont
  {{Watanabe-Nakayama}}}, \bibinfo {author} {\bibfnamefont {T.}~\bibnamefont
  {Okamoto}}, \bibinfo {author} {\bibfnamefont {S.}~\bibnamefont {Kudo}}, \
  and\ \bibinfo {author} {\bibfnamefont {E.}~\bibnamefont {Muneyuki}},\
  }\href@noop {} {\bibfield  {journal} {\bibinfo  {journal} {Proceedings of the
  National Academy of Sciences}\ }\textbf {\bibinfo {volume} {108}},\ \bibinfo
  {pages} {17951} (\bibinfo {year} {2011})}\BibitemShut {NoStop}%
\bibitem [{\citenamefont {Fletcher}\ and\ \citenamefont
  {Mullins}(2010)}]{fletcherCellMechanicsCytoskeleton2010}%
  \BibitemOpen
  \bibfield  {author} {\bibinfo {author} {\bibfnamefont {D.~A.}\ \bibnamefont
  {Fletcher}}\ and\ \bibinfo {author} {\bibfnamefont {R.~D.}\ \bibnamefont
  {Mullins}},\ }\href@noop {} {\bibfield  {journal} {\bibinfo  {journal}
  {Nature}\ }\textbf {\bibinfo {volume} {463}},\ \bibinfo {pages} {485}
  (\bibinfo {year} {2010})}\BibitemShut {NoStop}%
\bibitem [{\citenamefont {Yamada}\ and\ \citenamefont
  {Sixt}(2019)}]{yamadaMechanisms3DCell2019}%
  \BibitemOpen
  \bibfield  {author} {\bibinfo {author} {\bibfnamefont {K.~M.}\ \bibnamefont
  {Yamada}}\ and\ \bibinfo {author} {\bibfnamefont {M.}~\bibnamefont {Sixt}},\
  }\href@noop {} {\bibfield  {journal} {\bibinfo  {journal} {Nature Reviews
  Molecular Cell Biology}\ }\textbf {\bibinfo {volume} {20}},\ \bibinfo {pages}
  {738} (\bibinfo {year} {2019})}\BibitemShut {NoStop}%
\bibitem [{\citenamefont {Klein}\ and\ \citenamefont
  {Simons}(2011)}]{kleinUniversalPatternsStem2011a}%
  \BibitemOpen
  \bibfield  {author} {\bibinfo {author} {\bibfnamefont {A.~M.}\ \bibnamefont
  {Klein}}\ and\ \bibinfo {author} {\bibfnamefont {B.~D.}\ \bibnamefont
  {Simons}},\ }\href@noop {} {\bibfield  {journal} {\bibinfo  {journal}
  {Development}\ }\textbf {\bibinfo {volume} {138}},\ \bibinfo {pages} {3103}
  (\bibinfo {year} {2011})}\BibitemShut {NoStop}%
\bibitem [{\citenamefont {Barresi}\ and\ \citenamefont
  {Gilbert}(2019)}]{barresiDevelopmentalBiology2019}%
  \BibitemOpen
  \bibfield  {author} {\bibinfo {author} {\bibfnamefont {M.~J.~F.}\
  \bibnamefont {Barresi}}\ and\ \bibinfo {author} {\bibfnamefont {S.~F.}\
  \bibnamefont {Gilbert}},\ }\href@noop {} {\emph {\bibinfo {title}
  {Developmental {{Biology}}}}},\ \bibinfo {edition} {12th}\ ed.\ (\bibinfo
  {year} {2019})\BibitemShut {NoStop}%
\bibitem [{\citenamefont {Mammoto}\ and\ \citenamefont
  {Ingber}(2010)}]{mammotoMechanicalControlTissue2010a}%
  \BibitemOpen
  \bibfield  {author} {\bibinfo {author} {\bibfnamefont {T.}~\bibnamefont
  {Mammoto}}\ and\ \bibinfo {author} {\bibfnamefont {D.~E.}\ \bibnamefont
  {Ingber}},\ }\href@noop {} {\bibfield  {journal} {\bibinfo  {journal}
  {Development}\ }\textbf {\bibinfo {volume} {137}},\ \bibinfo {pages} {1407}
  (\bibinfo {year} {2010})}\BibitemShut {NoStop}%
\bibitem [{\citenamefont {Shawky}\ and\ \citenamefont
  {Davidson}(2015)}]{shawkyTissueMechanicsAdhesion2015}%
  \BibitemOpen
  \bibfield  {author} {\bibinfo {author} {\bibfnamefont {J.~H.}\ \bibnamefont
  {Shawky}}\ and\ \bibinfo {author} {\bibfnamefont {L.~A.}\ \bibnamefont
  {Davidson}},\ }\href@noop {} {\bibfield  {journal} {\bibinfo  {journal}
  {Developmental Biology}\ }\textbf {\bibinfo {volume} {401}},\ \bibinfo
  {pages} {152} (\bibinfo {year} {2015})}\BibitemShut {NoStop}%
\bibitem [{\citenamefont {Petridou}\ and\ \citenamefont
  {Heisenberg}(2019)}]{petridouTissueRheologyEmbryonic2019}%
  \BibitemOpen
  \bibfield  {author} {\bibinfo {author} {\bibfnamefont {N.~I.}\ \bibnamefont
  {Petridou}}\ and\ \bibinfo {author} {\bibfnamefont {C.-P.}\ \bibnamefont
  {Heisenberg}},\ }\href@noop {} {\bibfield  {journal} {\bibinfo  {journal}
  {The EMBO Journal}\ }\textbf {\bibinfo {volume} {38}},\ \bibinfo {pages}
  {e102497} (\bibinfo {year} {2019})}\BibitemShut {NoStop}%
\bibitem [{\citenamefont {Park}\ \emph {et~al.}(2015)\citenamefont {Park},
  \citenamefont {Kim}, \citenamefont {Bi}, \citenamefont {Mitchel},
  \citenamefont {Qazvini}, \citenamefont {Tantisira}, \citenamefont {Park},
  \citenamefont {McGill}, \citenamefont {Kim}, \citenamefont {Gweon},
  \citenamefont {Notbohm}, \citenamefont {Steward~Jr}, \citenamefont {Burger},
  \citenamefont {Randell}, \citenamefont {Kho}, \citenamefont {Tambe},
  \citenamefont {Hardin}, \citenamefont {Shore}, \citenamefont {Israel},
  \citenamefont {Weitz}, \citenamefont {Tschumperlin}, \citenamefont {Henske},
  \citenamefont {Weiss}, \citenamefont {Manning}, \citenamefont {Butler},
  \citenamefont {Drazen},\ and\ \citenamefont
  {Fredberg}}]{parkUnjammingCellShape2015}%
  \BibitemOpen
  \bibfield  {author} {\bibinfo {author} {\bibfnamefont {J.-A.}\ \bibnamefont
  {Park}}, \bibinfo {author} {\bibfnamefont {J.~H.}\ \bibnamefont {Kim}},
  \bibinfo {author} {\bibfnamefont {D.}~\bibnamefont {Bi}}, \bibinfo {author}
  {\bibfnamefont {J.~A.}\ \bibnamefont {Mitchel}}, \bibinfo {author}
  {\bibfnamefont {N.~T.}\ \bibnamefont {Qazvini}}, \bibinfo {author}
  {\bibfnamefont {K.}~\bibnamefont {Tantisira}}, \bibinfo {author}
  {\bibfnamefont {C.~Y.}\ \bibnamefont {Park}}, \bibinfo {author}
  {\bibfnamefont {M.}~\bibnamefont {McGill}}, \bibinfo {author} {\bibfnamefont
  {S.-H.}\ \bibnamefont {Kim}}, \bibinfo {author} {\bibfnamefont
  {B.}~\bibnamefont {Gweon}}, \bibinfo {author} {\bibfnamefont
  {J.}~\bibnamefont {Notbohm}}, \bibinfo {author} {\bibfnamefont
  {R.}~\bibnamefont {Steward~Jr}}, \bibinfo {author} {\bibfnamefont
  {S.}~\bibnamefont {Burger}}, \bibinfo {author} {\bibfnamefont {S.~H.}\
  \bibnamefont {Randell}}, \bibinfo {author} {\bibfnamefont {A.~T.}\
  \bibnamefont {Kho}}, \bibinfo {author} {\bibfnamefont {D.~T.}\ \bibnamefont
  {Tambe}}, \bibinfo {author} {\bibfnamefont {C.}~\bibnamefont {Hardin}},
  \bibinfo {author} {\bibfnamefont {S.~A.}\ \bibnamefont {Shore}}, \bibinfo
  {author} {\bibfnamefont {E.}~\bibnamefont {Israel}}, \bibinfo {author}
  {\bibfnamefont {D.~A.}\ \bibnamefont {Weitz}}, \bibinfo {author}
  {\bibfnamefont {D.~J.}\ \bibnamefont {Tschumperlin}}, \bibinfo {author}
  {\bibfnamefont {E.~P.}\ \bibnamefont {Henske}}, \bibinfo {author}
  {\bibfnamefont {S.~T.}\ \bibnamefont {Weiss}}, \bibinfo {author}
  {\bibfnamefont {M.~L.}\ \bibnamefont {Manning}}, \bibinfo {author}
  {\bibfnamefont {J.~P.}\ \bibnamefont {Butler}}, \bibinfo {author}
  {\bibfnamefont {J.~M.}\ \bibnamefont {Drazen}}, \ and\ \bibinfo {author}
  {\bibfnamefont {J.~J.}\ \bibnamefont {Fredberg}},\ }\href@noop {} {\bibfield
  {journal} {\bibinfo  {journal} {Nature Materials}\ }\textbf {\bibinfo
  {volume} {14}},\ \bibinfo {pages} {1040} (\bibinfo {year}
  {2015})}\BibitemShut {NoStop}%
\bibitem [{\citenamefont {Fodor}\ \emph {et~al.}(2018)\citenamefont {Fodor},
  \citenamefont {Mehandia}, \citenamefont {Comelles}, \citenamefont
  {Thiagarajan}, \citenamefont {Gov}, \citenamefont {Visco}, \citenamefont {van
  Wijland},\ and\ \citenamefont
  {Riveline}}]{fodorSpatialFluctuationsVertices2018a}%
  \BibitemOpen
  \bibfield  {author} {\bibinfo {author} {\bibfnamefont {{\'E}.}~\bibnamefont
  {Fodor}}, \bibinfo {author} {\bibfnamefont {V.}~\bibnamefont {Mehandia}},
  \bibinfo {author} {\bibfnamefont {J.}~\bibnamefont {Comelles}}, \bibinfo
  {author} {\bibfnamefont {R.}~\bibnamefont {Thiagarajan}}, \bibinfo {author}
  {\bibfnamefont {N.~S.}\ \bibnamefont {Gov}}, \bibinfo {author} {\bibfnamefont
  {P.}~\bibnamefont {Visco}}, \bibinfo {author} {\bibfnamefont
  {F.}~\bibnamefont {van Wijland}}, \ and\ \bibinfo {author} {\bibfnamefont
  {D.}~\bibnamefont {Riveline}},\ }\href@noop {} {\bibfield  {journal}
  {\bibinfo  {journal} {Biophysical Journal}\ }\textbf {\bibinfo {volume}
  {114}},\ \bibinfo {pages} {939} (\bibinfo {year} {2018})}\BibitemShut
  {NoStop}%
\bibitem [{\citenamefont {Mongera}\ \emph {et~al.}(2018)\citenamefont
  {Mongera}, \citenamefont {Rowghanian}, \citenamefont {Gustafson},
  \citenamefont {Shelton}, \citenamefont {Kealhofer}, \citenamefont {Carn},
  \citenamefont {Serwane}, \citenamefont {Lucio}, \citenamefont {Giammona},\
  and\ \citenamefont {Camp{\`a}s}}]{mongeraFluidtosolidJammingTransition2018}%
  \BibitemOpen
  \bibfield  {author} {\bibinfo {author} {\bibfnamefont {A.}~\bibnamefont
  {Mongera}}, \bibinfo {author} {\bibfnamefont {P.}~\bibnamefont {Rowghanian}},
  \bibinfo {author} {\bibfnamefont {H.~J.}\ \bibnamefont {Gustafson}}, \bibinfo
  {author} {\bibfnamefont {E.}~\bibnamefont {Shelton}}, \bibinfo {author}
  {\bibfnamefont {D.~A.}\ \bibnamefont {Kealhofer}}, \bibinfo {author}
  {\bibfnamefont {E.~K.}\ \bibnamefont {Carn}}, \bibinfo {author}
  {\bibfnamefont {F.}~\bibnamefont {Serwane}}, \bibinfo {author} {\bibfnamefont
  {A.~A.}\ \bibnamefont {Lucio}}, \bibinfo {author} {\bibfnamefont
  {J.}~\bibnamefont {Giammona}}, \ and\ \bibinfo {author} {\bibfnamefont
  {O.}~\bibnamefont {Camp{\`a}s}},\ }\href@noop {} {\bibfield  {journal}
  {\bibinfo  {journal} {Nature}\ }\textbf {\bibinfo {volume} {561}},\ \bibinfo
  {pages} {401} (\bibinfo {year} {2018})}\BibitemShut {NoStop}%
\bibitem [{\citenamefont {Angelini}\ \emph {et~al.}(2011)\citenamefont
  {Angelini}, \citenamefont {Hannezo}, \citenamefont {Trepat}, \citenamefont
  {Marquez}, \citenamefont {Fredberg},\ and\ \citenamefont
  {Weitz}}]{angeliniGlasslikeDynamicsCollective2011b}%
  \BibitemOpen
  \bibfield  {author} {\bibinfo {author} {\bibfnamefont {T.~E.}\ \bibnamefont
  {Angelini}}, \bibinfo {author} {\bibfnamefont {E.}~\bibnamefont {Hannezo}},
  \bibinfo {author} {\bibfnamefont {X.}~\bibnamefont {Trepat}}, \bibinfo
  {author} {\bibfnamefont {M.}~\bibnamefont {Marquez}}, \bibinfo {author}
  {\bibfnamefont {J.~J.}\ \bibnamefont {Fredberg}}, \ and\ \bibinfo {author}
  {\bibfnamefont {D.~A.}\ \bibnamefont {Weitz}},\ }\href@noop {} {\bibfield
  {journal} {\bibinfo  {journal} {Proceedings of the National Academy of
  Sciences}\ }\textbf {\bibinfo {volume} {108}},\ \bibinfo {pages} {4714}
  (\bibinfo {year} {2011})}\BibitemShut {NoStop}%
\bibitem [{\citenamefont {Wang}\ \emph {et~al.}(2020)\citenamefont {Wang},
  \citenamefont {Merkel}, \citenamefont {Sutter}, \citenamefont
  {{Erdemci-Tandogan}}, \citenamefont {Manning},\ and\ \citenamefont
  {Kasza}}]{wangAnisotropyLinksCell2020}%
  \BibitemOpen
  \bibfield  {author} {\bibinfo {author} {\bibfnamefont {X.}~\bibnamefont
  {Wang}}, \bibinfo {author} {\bibfnamefont {M.}~\bibnamefont {Merkel}},
  \bibinfo {author} {\bibfnamefont {L.~B.}\ \bibnamefont {Sutter}}, \bibinfo
  {author} {\bibfnamefont {G.}~\bibnamefont {{Erdemci-Tandogan}}}, \bibinfo
  {author} {\bibfnamefont {M.~L.}\ \bibnamefont {Manning}}, \ and\ \bibinfo
  {author} {\bibfnamefont {K.~E.}\ \bibnamefont {Kasza}},\ }\href@noop {}
  {\bibfield  {journal} {\bibinfo  {journal} {Proceedings of the National
  Academy of Sciences}\ }\textbf {\bibinfo {volume} {117}},\ \bibinfo {pages}
  {13541} (\bibinfo {year} {2020})}\BibitemShut {NoStop}%
\bibitem [{\citenamefont {Sussman}\ \emph
  {et~al.}(2018{\natexlab{a}})\citenamefont {Sussman}, \citenamefont
  {Paoluzzi}, \citenamefont {Cristina~Marchetti},\ and\ \citenamefont
  {Lisa~Manning}}]{sussmanAnomalousGlassyDynamics2018a}%
  \BibitemOpen
  \bibfield  {author} {\bibinfo {author} {\bibfnamefont {D.~M.}\ \bibnamefont
  {Sussman}}, \bibinfo {author} {\bibfnamefont {M.}~\bibnamefont {Paoluzzi}},
  \bibinfo {author} {\bibfnamefont {M.}~\bibnamefont {Cristina~Marchetti}}, \
  and\ \bibinfo {author} {\bibfnamefont {M.}~\bibnamefont {Lisa~Manning}},\
  }\href@noop {} {\bibfield  {journal} {\bibinfo  {journal} {EPL (Europhysics
  Letters)}\ }\textbf {\bibinfo {volume} {121}},\ \bibinfo {pages} {36001}
  (\bibinfo {year} {2018}{\natexlab{a}})}\BibitemShut {NoStop}%
\bibitem [{\citenamefont {Krajnc}(2020)}]{krajncSolidFluidTransition2020}%
  \BibitemOpen
  \bibfield  {author} {\bibinfo {author} {\bibfnamefont {M.}~\bibnamefont
  {Krajnc}},\ }\href@noop {} {\bibfield  {journal} {\bibinfo  {journal} {Soft
  Matter}\ }\textbf {\bibinfo {volume} {16}},\ \bibinfo {pages} {3209}
  (\bibinfo {year} {2020})}\BibitemShut {NoStop}%
\bibitem [{\citenamefont {Kim}\ \emph {et~al.}(2020)\citenamefont {Kim},
  \citenamefont {Pochitaloff}, \citenamefont {{Georgina-Stooke-Vaughan}},\ and\
  \citenamefont {Camp{\`a}s}}]{kimEmbryonicTissuesActive2020}%
  \BibitemOpen
  \bibfield  {author} {\bibinfo {author} {\bibfnamefont {S.}~\bibnamefont
  {Kim}}, \bibinfo {author} {\bibfnamefont {M.}~\bibnamefont {Pochitaloff}},
  \bibinfo {author} {\bibnamefont {{Georgina-Stooke-Vaughan}}}, \ and\ \bibinfo
  {author} {\bibfnamefont {O.}~\bibnamefont {Camp{\`a}s}},\ }\href@noop {}
  {\bibfield  {journal} {\bibinfo  {journal} {bioRxiv: 2020.06.17.157909}\ }
  (\bibinfo {year} {2020})}\BibitemShut {NoStop}%
\bibitem [{\citenamefont {Bi}\ \emph {et~al.}(2016)\citenamefont {Bi},
  \citenamefont {Yang}, \citenamefont {Marchetti},\ and\ \citenamefont
  {Manning}}]{biMotilityDrivenGlassJamming2016a}%
  \BibitemOpen
  \bibfield  {author} {\bibinfo {author} {\bibfnamefont {D.}~\bibnamefont
  {Bi}}, \bibinfo {author} {\bibfnamefont {X.}~\bibnamefont {Yang}}, \bibinfo
  {author} {\bibfnamefont {M.~C.}\ \bibnamefont {Marchetti}}, \ and\ \bibinfo
  {author} {\bibfnamefont {M.~L.}\ \bibnamefont {Manning}},\ }\href@noop {}
  {\bibfield  {journal} {\bibinfo  {journal} {Physical Review X}\ }\textbf
  {\bibinfo {volume} {6}},\ \bibinfo {pages} {021011} (\bibinfo {year}
  {2016})}\BibitemShut {NoStop}%
\bibitem [{\citenamefont {Chiang}\ and\ \citenamefont
  {Marenduzzo}(2016)}]{chiangGlassTransitionsCellular2016}%
  \BibitemOpen
  \bibfield  {author} {\bibinfo {author} {\bibfnamefont {M.}~\bibnamefont
  {Chiang}}\ and\ \bibinfo {author} {\bibfnamefont {D.}~\bibnamefont
  {Marenduzzo}},\ }\href@noop {} {\bibfield  {journal} {\bibinfo  {journal}
  {EPL (Europhysics Letters)}\ }\textbf {\bibinfo {volume} {116}},\ \bibinfo
  {pages} {28009} (\bibinfo {year} {2016})}\BibitemShut {NoStop}%
\bibitem [{\citenamefont {Yan}\ and\ \citenamefont
  {Bi}(2019)}]{yanMulticellularRosettesDrive2019}%
  \BibitemOpen
  \bibfield  {author} {\bibinfo {author} {\bibfnamefont {L.}~\bibnamefont
  {Yan}}\ and\ \bibinfo {author} {\bibfnamefont {D.}~\bibnamefont {Bi}},\
  }\href@noop {} {\bibfield  {journal} {\bibinfo  {journal} {Physical Review
  X}\ }\textbf {\bibinfo {volume} {9}},\ \bibinfo {pages} {011029} (\bibinfo
  {year} {2019})}\BibitemShut {NoStop}%
\bibitem [{\citenamefont {Erdemci-Tandogan}\ and\ \citenamefont
  {Manning}(2021)}]{erdemci2021effect}%
  \BibitemOpen
  \bibfield  {author} {\bibinfo {author} {\bibfnamefont {G.}~\bibnamefont
  {Erdemci-Tandogan}}\ and\ \bibinfo {author} {\bibfnamefont {M.~L.}\
  \bibnamefont {Manning}},\ }\href@noop {} {\bibfield  {journal} {\bibinfo
  {journal} {PLOS Computational Biology}\ }\textbf {\bibinfo {volume} {17}},\
  \bibinfo {pages} {e1009049} (\bibinfo {year} {2021})}\BibitemShut {NoStop}%
\bibitem [{\citenamefont {Das}\ \emph {et~al.}(2020)\citenamefont {Das},
  \citenamefont {Sastry},\ and\ \citenamefont
  {Bi}}]{dasControlledNeighborExchanges2020a}%
  \BibitemOpen
  \bibfield  {author} {\bibinfo {author} {\bibfnamefont {A.}~\bibnamefont
  {Das}}, \bibinfo {author} {\bibfnamefont {S.}~\bibnamefont {Sastry}}, \ and\
  \bibinfo {author} {\bibfnamefont {D.}~\bibnamefont {Bi}},\ }\href@noop {}
  {\bibfield  {journal} {\bibinfo  {journal} {arXiv:2003.01042 [cond-mat,
  q-bio]}\ } (\bibinfo {year} {2020})}\BibitemShut {NoStop}%
\bibitem [{\citenamefont {Harding}\ \emph {et~al.}(2014)\citenamefont
  {Harding}, \citenamefont {McGraw},\ and\ \citenamefont
  {Nechiporuk}}]{hardingRolesRegulationMulticellular2014a}%
  \BibitemOpen
  \bibfield  {author} {\bibinfo {author} {\bibfnamefont {M.~J.}\ \bibnamefont
  {Harding}}, \bibinfo {author} {\bibfnamefont {H.~F.}\ \bibnamefont {McGraw}},
  \ and\ \bibinfo {author} {\bibfnamefont {A.}~\bibnamefont {Nechiporuk}},\
  }\href@noop {} {\bibfield  {journal} {\bibinfo  {journal} {Development}\
  }\textbf {\bibinfo {volume} {141}},\ \bibinfo {pages} {2549} (\bibinfo {year}
  {2014})}\BibitemShut {NoStop}%
\bibitem [{\citenamefont {Trichas}\ \emph {et~al.}(2012)\citenamefont
  {Trichas}, \citenamefont {Smith}, \citenamefont {White}, \citenamefont
  {Wilkins}, \citenamefont {Watanabe}, \citenamefont {Moore}, \citenamefont
  {Joyce}, \citenamefont {Sugnaseelan}, \citenamefont {Rodriguez},
  \citenamefont {Kay}, \citenamefont {Baker}, \citenamefont {Maini},\ and\
  \citenamefont {Srinivas}}]{trichasMultiCellularRosettesMouse2012a}%
  \BibitemOpen
  \bibfield  {author} {\bibinfo {author} {\bibfnamefont {G.}~\bibnamefont
  {Trichas}}, \bibinfo {author} {\bibfnamefont {A.~M.}\ \bibnamefont {Smith}},
  \bibinfo {author} {\bibfnamefont {N.}~\bibnamefont {White}}, \bibinfo
  {author} {\bibfnamefont {V.}~\bibnamefont {Wilkins}}, \bibinfo {author}
  {\bibfnamefont {T.}~\bibnamefont {Watanabe}}, \bibinfo {author}
  {\bibfnamefont {A.}~\bibnamefont {Moore}}, \bibinfo {author} {\bibfnamefont
  {B.}~\bibnamefont {Joyce}}, \bibinfo {author} {\bibfnamefont
  {J.}~\bibnamefont {Sugnaseelan}}, \bibinfo {author} {\bibfnamefont {T.~A.}\
  \bibnamefont {Rodriguez}}, \bibinfo {author} {\bibfnamefont {D.}~\bibnamefont
  {Kay}}, \bibinfo {author} {\bibfnamefont {R.~E.}\ \bibnamefont {Baker}},
  \bibinfo {author} {\bibfnamefont {P.~K.}\ \bibnamefont {Maini}}, \ and\
  \bibinfo {author} {\bibfnamefont {S.}~\bibnamefont {Srinivas}},\ }\href@noop
  {} {\bibfield  {journal} {\bibinfo  {journal} {PLOS Biology}\ }\textbf
  {\bibinfo {volume} {10}},\ \bibinfo {pages} {e1001256} (\bibinfo {year}
  {2012})}\BibitemShut {NoStop}%
\bibitem [{\citenamefont {Blankenship}\ \emph {et~al.}(2006)\citenamefont
  {Blankenship}, \citenamefont {Backovic}, \citenamefont {Sanny}, \citenamefont
  {Weitz},\ and\ \citenamefont
  {Zallen}}]{blankenshipMulticellularRosetteFormation2006}%
  \BibitemOpen
  \bibfield  {author} {\bibinfo {author} {\bibfnamefont {J.~T.}\ \bibnamefont
  {Blankenship}}, \bibinfo {author} {\bibfnamefont {S.~T.}\ \bibnamefont
  {Backovic}}, \bibinfo {author} {\bibfnamefont {J.~S.~P.}\ \bibnamefont
  {Sanny}}, \bibinfo {author} {\bibfnamefont {O.}~\bibnamefont {Weitz}}, \ and\
  \bibinfo {author} {\bibfnamefont {J.~A.}\ \bibnamefont {Zallen}},\
  }\href@noop {} {\bibfield  {journal} {\bibinfo  {journal} {Developmental
  Cell}\ }\textbf {\bibinfo {volume} {11}},\ \bibinfo {pages} {459} (\bibinfo
  {year} {2006})}\BibitemShut {NoStop}%
\bibitem [{\citenamefont {Berthier}\ \emph {et~al.}(2017)\citenamefont
  {Berthier}, \citenamefont {Flenner},\ and\ \citenamefont
  {Szamel}}]{berthierHowActiveForces2017}%
  \BibitemOpen
  \bibfield  {author} {\bibinfo {author} {\bibfnamefont {L.}~\bibnamefont
  {Berthier}}, \bibinfo {author} {\bibfnamefont {E.}~\bibnamefont {Flenner}}, \
  and\ \bibinfo {author} {\bibfnamefont {G.}~\bibnamefont {Szamel}},\ }\href
  {\doibase 10.1088/1367-2630/aa914e} {\bibfield  {journal} {\bibinfo
  {journal} {New Journal of Physics}\ }\textbf {\bibinfo {volume} {19}},\
  \bibinfo {pages} {125006} (\bibinfo {year} {2017})}\BibitemShut {NoStop}%
\bibitem [{\citenamefont {Curran}\ \emph {et~al.}(2017)\citenamefont {Curran},
  \citenamefont {Strandkvist}, \citenamefont {Bathmann}, \citenamefont {{de
  Gennes}}, \citenamefont {Kabla}, \citenamefont {Salbreux},\ and\
  \citenamefont {Baum}}]{curranMyosinIIControls2017}%
  \BibitemOpen
  \bibfield  {author} {\bibinfo {author} {\bibfnamefont {S.}~\bibnamefont
  {Curran}}, \bibinfo {author} {\bibfnamefont {C.}~\bibnamefont {Strandkvist}},
  \bibinfo {author} {\bibfnamefont {J.}~\bibnamefont {Bathmann}}, \bibinfo
  {author} {\bibfnamefont {M.}~\bibnamefont {{de Gennes}}}, \bibinfo {author}
  {\bibfnamefont {A.}~\bibnamefont {Kabla}}, \bibinfo {author} {\bibfnamefont
  {G.}~\bibnamefont {Salbreux}}, \ and\ \bibinfo {author} {\bibfnamefont
  {B.}~\bibnamefont {Baum}},\ }\href@noop {} {\bibfield  {journal} {\bibinfo
  {journal} {Developmental Cell}\ }\textbf {\bibinfo {volume} {43}},\ \bibinfo
  {pages} {1} (\bibinfo {year} {2017})}\BibitemShut {NoStop}%
\bibitem [{\citenamefont {Tetley}\ \emph {et~al.}(2019)\citenamefont {Tetley},
  \citenamefont {Staddon}, \citenamefont {Heller}, \citenamefont {Hoppe},
  \citenamefont {Banerjee},\ and\ \citenamefont
  {Mao}}]{tetleyTissueFluidityPromotes2019}%
  \BibitemOpen
  \bibfield  {author} {\bibinfo {author} {\bibfnamefont {R.~J.}\ \bibnamefont
  {Tetley}}, \bibinfo {author} {\bibfnamefont {M.~F.}\ \bibnamefont {Staddon}},
  \bibinfo {author} {\bibfnamefont {D.}~\bibnamefont {Heller}}, \bibinfo
  {author} {\bibfnamefont {A.}~\bibnamefont {Hoppe}}, \bibinfo {author}
  {\bibfnamefont {S.}~\bibnamefont {Banerjee}}, \ and\ \bibinfo {author}
  {\bibfnamefont {Y.}~\bibnamefont {Mao}},\ }\href {\doibase
  10.1038/s41567-019-0618-1} {\bibfield  {journal} {\bibinfo  {journal} {Nature
  Physics}\ }\textbf {\bibinfo {volume} {15}},\ \bibinfo {pages} {1195}
  (\bibinfo {year} {2019})}\BibitemShut {NoStop}%
\bibitem [{\citenamefont {Honda}\ \emph {et~al.}(1984)\citenamefont {Honda},
  \citenamefont {Yamanaka},\ and\ \citenamefont
  {{Dan-Sohkawa}}}]{hondaComputerSimulationGeometrical1984}%
  \BibitemOpen
  \bibfield  {author} {\bibinfo {author} {\bibfnamefont {H.}~\bibnamefont
  {Honda}}, \bibinfo {author} {\bibfnamefont {H.}~\bibnamefont {Yamanaka}}, \
  and\ \bibinfo {author} {\bibfnamefont {M.}~\bibnamefont {{Dan-Sohkawa}}},\
  }\href@noop {} {\bibfield  {journal} {\bibinfo  {journal} {Journal of
  Theoretical Biology}\ }\textbf {\bibinfo {volume} {106}},\ \bibinfo {pages}
  {423} (\bibinfo {year} {1984})}\BibitemShut {NoStop}%
\bibitem [{\citenamefont {Bosveld}\ \emph {et~al.}(2018)\citenamefont
  {Bosveld}, \citenamefont {Wang},\ and\ \citenamefont
  {Bella{\"i}che}}]{bosveldTricellularJunctionsHot2018}%
  \BibitemOpen
  \bibfield  {author} {\bibinfo {author} {\bibfnamefont {F.}~\bibnamefont
  {Bosveld}}, \bibinfo {author} {\bibfnamefont {Z.}~\bibnamefont {Wang}}, \
  and\ \bibinfo {author} {\bibfnamefont {Y.}~\bibnamefont {Bella{\"i}che}},\
  }\href {\doibase 10.1016/j.ceb.2018.05.002} {\bibfield  {journal} {\bibinfo
  {journal} {Current Opinion in Cell Biology}\ }\bibinfo {series} {Cell
  {{Dynamics}}},\ \textbf {\bibinfo {volume} {54}},\ \bibinfo {pages} {80}
  (\bibinfo {year} {2018})}\BibitemShut {NoStop}%
\bibitem [{\citenamefont {Bi}\ \emph {et~al.}(2015)\citenamefont {Bi},
  \citenamefont {Lopez}, \citenamefont {Schwarz},\ and\ \citenamefont
  {Manning}}]{biDensityindependentRigidityTransition2015a}%
  \BibitemOpen
  \bibfield  {author} {\bibinfo {author} {\bibfnamefont {D.}~\bibnamefont
  {Bi}}, \bibinfo {author} {\bibfnamefont {J.~H.}\ \bibnamefont {Lopez}},
  \bibinfo {author} {\bibfnamefont {J.~M.}\ \bibnamefont {Schwarz}}, \ and\
  \bibinfo {author} {\bibfnamefont {M.~L.}\ \bibnamefont {Manning}},\
  }\href@noop {} {\bibfield  {journal} {\bibinfo  {journal} {Nature Physics}\
  }\textbf {\bibinfo {volume} {11}},\ \bibinfo {pages} {1074} (\bibinfo {year}
  {2015})}\BibitemShut {NoStop}%
\bibitem [{\citenamefont {Merkel}\ \emph {et~al.}(2019)\citenamefont {Merkel},
  \citenamefont {Baumgarten}, \citenamefont {Tighe},\ and\ \citenamefont
  {Manning}}]{merkelMinimallengthApproachUnifies2019}%
  \BibitemOpen
  \bibfield  {author} {\bibinfo {author} {\bibfnamefont {M.}~\bibnamefont
  {Merkel}}, \bibinfo {author} {\bibfnamefont {K.}~\bibnamefont {Baumgarten}},
  \bibinfo {author} {\bibfnamefont {B.~P.}\ \bibnamefont {Tighe}}, \ and\
  \bibinfo {author} {\bibfnamefont {M.~L.}\ \bibnamefont {Manning}},\ }\href
  {\doibase 10.1073/pnas.1815436116} {\bibfield  {journal} {\bibinfo  {journal}
  {Proceedings of the National Academy of Sciences}\ }\textbf {\bibinfo
  {volume} {116}},\ \bibinfo {pages} {6560} (\bibinfo {year}
  {2019})}\BibitemShut {NoStop}%
\bibitem [{\citenamefont {Czajkowski}\ \emph {et~al.}(2019)\citenamefont
  {Czajkowski}, \citenamefont {Sussman}, \citenamefont {Marchetti},\ and\
  \citenamefont {Manning}}]{czajkowskiGlassyDynamicsModels2019a}%
  \BibitemOpen
  \bibfield  {author} {\bibinfo {author} {\bibfnamefont {M.}~\bibnamefont
  {Czajkowski}}, \bibinfo {author} {\bibfnamefont {D.~M.}\ \bibnamefont
  {Sussman}}, \bibinfo {author} {\bibfnamefont {M.~C.}\ \bibnamefont
  {Marchetti}}, \ and\ \bibinfo {author} {\bibfnamefont {M.~L.}\ \bibnamefont
  {Manning}},\ }\href {\doibase 10.1039/C9SM00916G} {\bibfield  {journal}
  {\bibinfo  {journal} {Soft Matter}\ }\textbf {\bibinfo {volume} {15}},\
  \bibinfo {pages} {9133} (\bibinfo {year} {2019})}\BibitemShut {NoStop}%
\bibitem [{\citenamefont {Bodrova}\ \emph {et~al.}(2016)\citenamefont
  {Bodrova}, \citenamefont {Chechkin}, \citenamefont {Cherstvy}, \citenamefont
  {Safdari}, \citenamefont {Sokolov},\ and\ \citenamefont
  {Metzler}}]{bodrovaUnderdampedScaledBrownian2016}%
  \BibitemOpen
  \bibfield  {author} {\bibinfo {author} {\bibfnamefont {A.~S.}\ \bibnamefont
  {Bodrova}}, \bibinfo {author} {\bibfnamefont {A.~V.}\ \bibnamefont
  {Chechkin}}, \bibinfo {author} {\bibfnamefont {A.~G.}\ \bibnamefont
  {Cherstvy}}, \bibinfo {author} {\bibfnamefont {H.}~\bibnamefont {Safdari}},
  \bibinfo {author} {\bibfnamefont {I.~M.}\ \bibnamefont {Sokolov}}, \ and\
  \bibinfo {author} {\bibfnamefont {R.}~\bibnamefont {Metzler}},\ }\href@noop
  {} {\bibfield  {journal} {\bibinfo  {journal} {Scientific Reports}\ }\textbf
  {\bibinfo {volume} {6}},\ \bibinfo {pages} {30520} (\bibinfo {year}
  {2016})}\BibitemShut {NoStop}%
\bibitem [{\citenamefont {Nicolas}\ \emph {et~al.}(2018)\citenamefont
  {Nicolas}, \citenamefont {Ferrero}, \citenamefont {Martens},\ and\
  \citenamefont {Barrat}}]{nicolasDeformationFlowAmorphous2018}%
  \BibitemOpen
  \bibfield  {author} {\bibinfo {author} {\bibfnamefont {A.}~\bibnamefont
  {Nicolas}}, \bibinfo {author} {\bibfnamefont {E.~E.}\ \bibnamefont
  {Ferrero}}, \bibinfo {author} {\bibfnamefont {K.}~\bibnamefont {Martens}}, \
  and\ \bibinfo {author} {\bibfnamefont {J.-L.}\ \bibnamefont {Barrat}},\
  }\href {\doibase 10.1103/RevModPhys.90.045006} {\bibfield  {journal}
  {\bibinfo  {journal} {Reviews of Modern Physics}\ }\textbf {\bibinfo {volume}
  {90}},\ \bibinfo {pages} {045006} (\bibinfo {year} {2018})}\BibitemShut
  {NoStop}%
\bibitem [{\citenamefont {Debets}\ \emph {et~al.}(2021)\citenamefont {Debets},
  \citenamefont {{de Wit}},\ and\ \citenamefont
  {Janssen}}]{debetsCageLengthControls2021}%
  \BibitemOpen
  \bibfield  {author} {\bibinfo {author} {\bibfnamefont {V.~E.}\ \bibnamefont
  {Debets}}, \bibinfo {author} {\bibfnamefont {X.~M.}\ \bibnamefont {{de
  Wit}}}, \ and\ \bibinfo {author} {\bibfnamefont {L.~M.~C.}\ \bibnamefont
  {Janssen}},\ }\href@noop {} {\bibfield  {journal} {\bibinfo  {journal}
  {arXiv:2111.11171}\ } (\bibinfo {year} {2021})}\BibitemShut {NoStop}%
\bibitem [{\citenamefont {Kasza}\ \emph {et~al.}(2014)\citenamefont {Kasza},
  \citenamefont {Farrell},\ and\ \citenamefont
  {Zallen}}]{kaszaSpatiotemporalControlEpithelial2014a}%
  \BibitemOpen
  \bibfield  {author} {\bibinfo {author} {\bibfnamefont {K.~E.}\ \bibnamefont
  {Kasza}}, \bibinfo {author} {\bibfnamefont {D.~L.}\ \bibnamefont {Farrell}},
  \ and\ \bibinfo {author} {\bibfnamefont {J.~A.}\ \bibnamefont {Zallen}},\
  }\href@noop {} {\bibfield  {journal} {\bibinfo  {journal} {Proceedings of the
  National Academy of Sciences}\ }\textbf {\bibinfo {volume} {111}},\ \bibinfo
  {pages} {11732} (\bibinfo {year} {2014})}\BibitemShut {NoStop}%
\bibitem [{\citenamefont {Sussman}\ \emph
  {et~al.}(2018{\natexlab{b}})\citenamefont {Sussman}, \citenamefont {Schwarz},
  \citenamefont {Marchetti},\ and\ \citenamefont
  {Manning}}]{sussmanSoftSharpInterfaces2018b}%
  \BibitemOpen
  \bibfield  {author} {\bibinfo {author} {\bibfnamefont {D.~M.}\ \bibnamefont
  {Sussman}}, \bibinfo {author} {\bibfnamefont {J.~M.}\ \bibnamefont
  {Schwarz}}, \bibinfo {author} {\bibfnamefont {M.~C.}\ \bibnamefont
  {Marchetti}}, \ and\ \bibinfo {author} {\bibfnamefont {M.~L.}\ \bibnamefont
  {Manning}},\ }\href@noop {} {\bibfield  {journal} {\bibinfo  {journal}
  {Physical Review Letters}\ }\textbf {\bibinfo {volume} {120}},\ \bibinfo
  {pages} {058001} (\bibinfo {year} {2018}{\natexlab{b}})}\BibitemShut
  {NoStop}%
\end{thebibliography}%


\begin{thebibliography}{4}%
\makeatletter
\providecommand \@ifxundefined [1]{%
 \@ifx{#1\undefined}
}%
\providecommand \@ifnum [1]{%
 \ifnum #1\expandafter \@firstoftwo
 \else \expandafter \@secondoftwo
 \fi
}%
\providecommand \@ifx [1]{%
 \ifx #1\expandafter \@firstoftwo
 \else \expandafter \@secondoftwo
 \fi
}%
\providecommand \natexlab [1]{#1}%
\providecommand \enquote  [1]{``#1''}%
\providecommand \bibnamefont  [1]{#1}%
\providecommand \bibfnamefont [1]{#1}%
\providecommand \citenamefont [1]{#1}%
\providecommand \href@noop [0]{\@secondoftwo}%
\providecommand \href [0]{\begingroup \@sanitize@url \@href}%
\providecommand \@href[1]{\@@startlink{#1}\@@href}%
\providecommand \@@href[1]{\endgroup#1\@@endlink}%
\providecommand \@sanitize@url [0]{\catcode `\\12\catcode `\$12\catcode
  `\&12\catcode `\#12\catcode `\^12\catcode `\_12\catcode `\%12\relax}%
\providecommand \@@startlink[1]{}%
\providecommand \@@endlink[0]{}%
\providecommand \url  [0]{\begingroup\@sanitize@url \@url }%
\providecommand \@url [1]{\endgroup\@href {#1}{\urlprefix }}%
\providecommand \urlprefix  [0]{URL }%
\providecommand \Eprint [0]{\href }%
\providecommand \doibase [0]{http://dx.doi.org/}%
\providecommand \selectlanguage [0]{\@gobble}%
\providecommand \bibinfo  [0]{\@secondoftwo}%
\providecommand \bibfield  [0]{\@secondoftwo}%
\providecommand \translation [1]{[#1]}%
\providecommand \BibitemOpen [0]{}%
\providecommand \bibitemStop [0]{}%
\providecommand \bibitemNoStop [0]{.\EOS\space}%
\providecommand \EOS [0]{\spacefactor3000\relax}%
\providecommand \BibitemShut  [1]{\csname bibitem#1\endcsname}%
\let\auto@bib@innerbib\@empty
\bibitem [{\citenamefont {Bodrova}\ \emph {et~al.}(2016)\citenamefont
  {Bodrova}, \citenamefont {Chechkin}, \citenamefont {Cherstvy}, \citenamefont
  {Safdari}, \citenamefont {Sokolov},\ and\ \citenamefont
  {Metzler}}]{bodrovaUnderdampedScaledBrownian2016}%
  \BibitemOpen
  \bibfield  {author} {\bibinfo {author} {\bibfnamefont {A.~S.}\ \bibnamefont
  {Bodrova}}, \bibinfo {author} {\bibfnamefont {A.~V.}\ \bibnamefont
  {Chechkin}}, \bibinfo {author} {\bibfnamefont {A.~G.}\ \bibnamefont
  {Cherstvy}}, \bibinfo {author} {\bibfnamefont {H.}~\bibnamefont {Safdari}},
  \bibinfo {author} {\bibfnamefont {I.~M.}\ \bibnamefont {Sokolov}}, \ and\
  \bibinfo {author} {\bibfnamefont {R.}~\bibnamefont {Metzler}},\ }\href@noop
  {} {\bibfield  {journal} {\bibinfo  {journal} {Scientific Reports}\ }\textbf
  {\bibinfo {volume} {6}},\ \bibinfo {pages} {30520} (\bibinfo {year}
  {2016})}\BibitemShut {NoStop}%
\bibitem [{\citenamefont {Yan}\ and\ \citenamefont
  {Bi}(2019)}]{yanMulticellularRosettesDrive2019}%
  \BibitemOpen
  \bibfield  {author} {\bibinfo {author} {\bibfnamefont {L.}~\bibnamefont
  {Yan}}\ and\ \bibinfo {author} {\bibfnamefont {D.}~\bibnamefont {Bi}},\
  }\href@noop {} {\bibfield  {journal} {\bibinfo  {journal} {Physical Review
  X}\ }\textbf {\bibinfo {volume} {9}},\ \bibinfo {pages} {011029} (\bibinfo
  {year} {2019})}\BibitemShut {NoStop}%
\bibitem [{\citenamefont {Spencer}\ \emph {et~al.}(2017)\citenamefont
  {Spencer}, \citenamefont {Jabeen},\ and\ \citenamefont
  {Lubensky}}]{spencerVertexStabilityTopological2017}%
  \BibitemOpen
  \bibfield  {author} {\bibinfo {author} {\bibfnamefont {M.~A.}\ \bibnamefont
  {Spencer}}, \bibinfo {author} {\bibfnamefont {Z.}~\bibnamefont {Jabeen}}, \
  and\ \bibinfo {author} {\bibfnamefont {D.~K.}\ \bibnamefont {Lubensky}},\
  }\href@noop {} {\bibfield  {journal} {\bibinfo  {journal} {The European
  Physical Journal E}\ }\textbf {\bibinfo {volume} {40}},\ \bibinfo {pages} {2}
  (\bibinfo {year} {2017})}\BibitemShut {NoStop}%
\bibitem [{\citenamefont {Sussman}\ \emph {et~al.}(2018)\citenamefont
  {Sussman}, \citenamefont {Schwarz}, \citenamefont {Marchetti},\ and\
  \citenamefont {Manning}}]{sussmanSoftSharpInterfaces2018b}%
  \BibitemOpen
  \bibfield  {author} {\bibinfo {author} {\bibfnamefont {D.~M.}\ \bibnamefont
  {Sussman}}, \bibinfo {author} {\bibfnamefont {J.~M.}\ \bibnamefont
  {Schwarz}}, \bibinfo {author} {\bibfnamefont {M.~C.}\ \bibnamefont
  {Marchetti}}, \ and\ \bibinfo {author} {\bibfnamefont {M.~L.}\ \bibnamefont
  {Manning}},\ }\href@noop {} {\bibfield  {journal} {\bibinfo  {journal}
  {Physical Review Letters}\ }\textbf {\bibinfo {volume} {120}},\ \bibinfo
  {pages} {058001} (\bibinfo {year} {2018})}\BibitemShut {NoStop}%
\end{thebibliography}%

\end{document}


\title{SUPPLEMENTAL MATERIAL\\
Non-monotonic fluidization generated by fluctuating edge tensions in confluent tissues}

\author{Takaki Yamamoto}
\email{takaki.yamamoto@riken.jp}
\affiliation{Laboratory for Physical Biology, RIKEN Center for Biosystems Dynamics Research, Kobe 650-0047, Japan}
\affiliation{Nonequilibrium Physics of Living Matter RIKEN Hakubi Research Team, RIKEN Center for Biosystems Dynamics Research, 2-2-3 Minatojima-minamimachi, Chuo-ku, Kobe 650-0047, Japan}
 
\author{Daniel M. Sussman}%
\affiliation{Department of Physics, Emory University, Atlanta, GA, USA}%
\author{Tatsuo Shibata}
\affiliation{Laboratory for Physical Biology, RIKEN Center for Biosystems Dynamics Research, Kobe 650-0047, Japan}

\author{M. Lisa Manning}%
 \email{memanning@gmail.com}
\affiliation{Department of Physics, Syracuse University, Syracuse, New York 13244, USA
}%
\affiliation{BioInspired Institute, Syracuse University, Syracuse, New York 13244, USA}%

\maketitle

\section{Time scales for edge shrinking and growing\label{appendix_FPT}}
As discussed in the main text, we expect that a cell center of mass diffuses on the timescale associated with either edges shrinking to zero and inducing a T1 transition or growing beyond unity and triggering a T1 transition in a nearby edge.  While in the main text we focus on the diffusion timescale of an edge in the absence of any boundary conditions, here we numerically study a toy first-passage-time problem in order to determine if the first-passage time statistics with absorbing boundary conditions are different from the simple diffusion problem.

Specifically, we are interested in the behavior of an edge subject to the dynamics given by Eq.~(4) and Eq.~(3) in the main text, and in calculating the time it takes for such an edge to either shrink to zero length (where the full model would attempt a T1 transition) or grows significantly longer than unity (where it is likely that nearby edges exhibit a T1) in the absence of other interactions. 

\begin{figure}[h]
 \begin{center}
  \includegraphics[width=85mm]{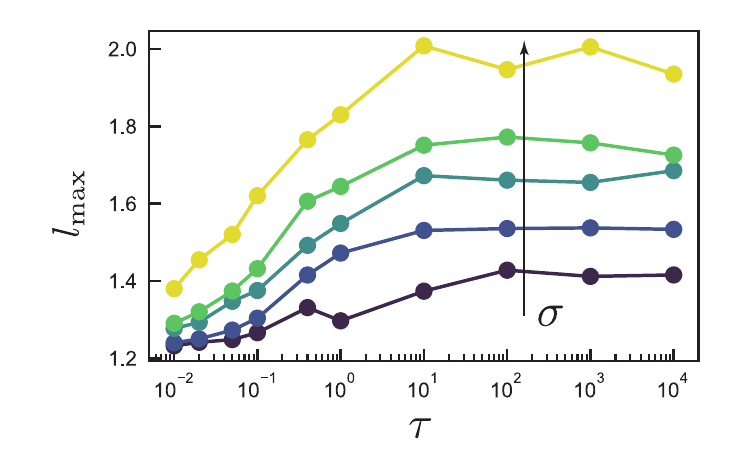}
 \end{center}
 \caption{\textbf{Maximum edge length $l_{\rm max}$.}
 $l_{\rm max}$ vs. $\tau$ for $\sigma \in [ 0.02, 0.05, 0.10,0.15, 0.30]$ (from dark color to light color). }
 \label{fig_lmax}
\end{figure}

To quantify this upper edge length cutoff, we analyze the edge-length distributions in the full numerical simulations, and Fig.~\ref{fig_lmax} shows the maximum value of \ty{edge} length in our finite simulation box, $l_{\rm max}$, as a function of model parameters. 
\ty{Since Fig.~\ref{fig_lmax} suggests that a T1 transition is induced to change the geometry when the edge length exceeds $\sim1.2\sim2l_{\rm hex}\ (l_{\rm hex}=\sqrt{2\sqrt{3}}/3\approx 0.62)$,
we use that value $l_{l}=2l_{\rm hex}$ as the upper length cutoff in our toy model. }

We numerically solve Eq.~(4) using Eq.~(3) with $l_{ij}=l$ and $\Delta\lambda_{ij}=\Delta\lambda$ with the initial edge length $l=l_{\rm hex}$ and then calculate the time $\tau_{\rm FPT}$ when the edge length satisfies $l<0$ or $l_{l}<l$ for the first time, which is a first passage time problem. 
We used forward Euler method with a time step $\delta t=0.01$.
$\Delta \lambda$ is sampled at the initial time point $t=0$ from the normal distribution $N(0,\sigma^2)$, which is the stationary distribution of Eq.~(3). 

\begin{figure}[h]
 \begin{center}
  \includegraphics[width=85mm]{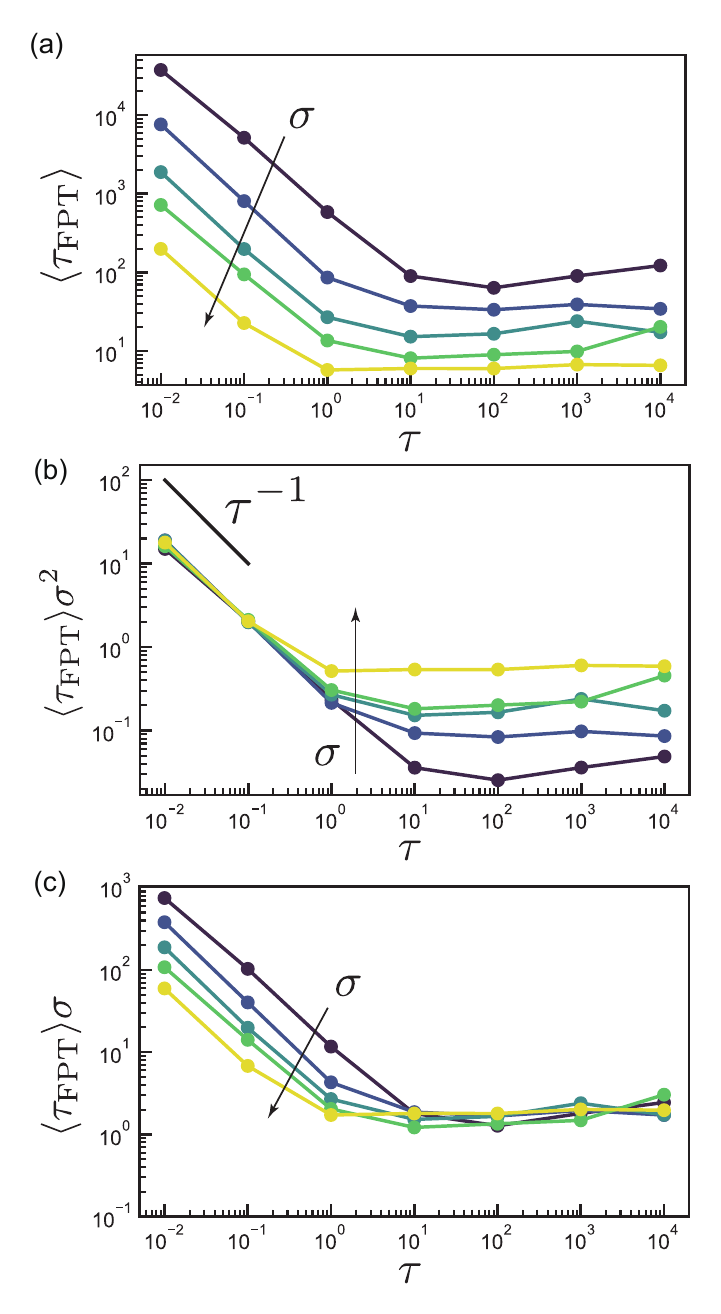}
 \end{center}
 \caption{\textbf{Scaling behavior of the average first passage time $\langle \tau_{\rm FPT}\rangle$ in the 2 vertex model.
 }(a) $\langle \tau_{\rm FPT}\rangle$ vs. $\tau$. (b) $\langle \tau_{\rm FPT}\rangle \sigma^2$ vs. $\tau$. The solid line is a guide for eyes indicating the power law $\tau^{-1}$. (c) $\langle \tau_{\rm FPT}\rangle \sigma$ vs. $\tau$. 
$\sigma \in [ 0.02, 0.05, 0.10,0.15, 0.30]$ (from dark color to light color). }
 \label{fig_FPT}
\end{figure}

Figure~\ref{fig_FPT}(a) shows the average first passage time $\langle \tau_{\rm FPT}\rangle$ calculated from $100$ trajectories for each set of parameters $(\sigma,\tau)$. 
For each simulation, we stopped calculation if the edge length keeps the condition $0<l<l_{l}$ within $10^5$ natural time units. 
We avoid such trajectories as rare events in calculating $\langle \tau_{\rm FPT}\rangle$\ty{; they were only observed for $(\sigma,\tau) = (0.02,0.01)$, where 9 of 100 trajectories did not satisfy $l<0$ or $l_{l}<l$ within $10^5$ natural time units.}
To investigate the scaling behavior in the small- and large-$\tau$ regimes, we also plot $\langle \tau_{\rm FPT}\rangle \sigma^2$ vs. $\tau$ and $\langle \tau_{\rm FPT}\rangle \sigma$ vs. $\tau$ in Fig.~\ref{fig_FPT}(b) and (c). 
Figure~\ref{fig_FPT}(b) and (c) suggest that $\langle \tau_{\rm FPT}\rangle \sim 1/ \sigma^2\tau$ and $\langle \tau_{\rm FPT}\rangle \sim 1/\sigma$ in the small- and large-$\tau$ regimes, respectively. 

As discussed in the main text, an analytical calculation of the diffusion of a single edge in the absence of any other interactions gives ${\rm MSD}_{l}(t) = 2\sigma^2\tau t + 2\sigma^2\tau^2(\exp(-t/\tau)-1)$, in which ${\rm MSD}_{l}(t)\sim 2\sigma^2\tau t$ ($t \gg \tau$) and ${\rm MSD}_{l}(t)\sim \sigma^2 t^2$ ($\tau \gg t$)\ty{~\cite{bodrovaUnderdampedScaledBrownian2016}}.

Therefore, in the small-$\tau$ regime, our numerical result for the first passage time, $\langle \tau_{\rm FPT}\rangle \approx l_{\rm hex}/ 2\sigma^2\tau$, exhibits the same scaling as the analytical MSD prediction.  Interestingly, the MSD scaling for the large $\tau$ regime is also consistent with the large $\tau$ FPT result: $\langle \tau_{\rm FPT}\rangle \approx l_{\rm hex}/\sigma$.

\section{Scaling of the diffusion constant for the resetting and resampling models}

\begin{figure}[h]
 \begin{center}
  \includegraphics[width=178mm]{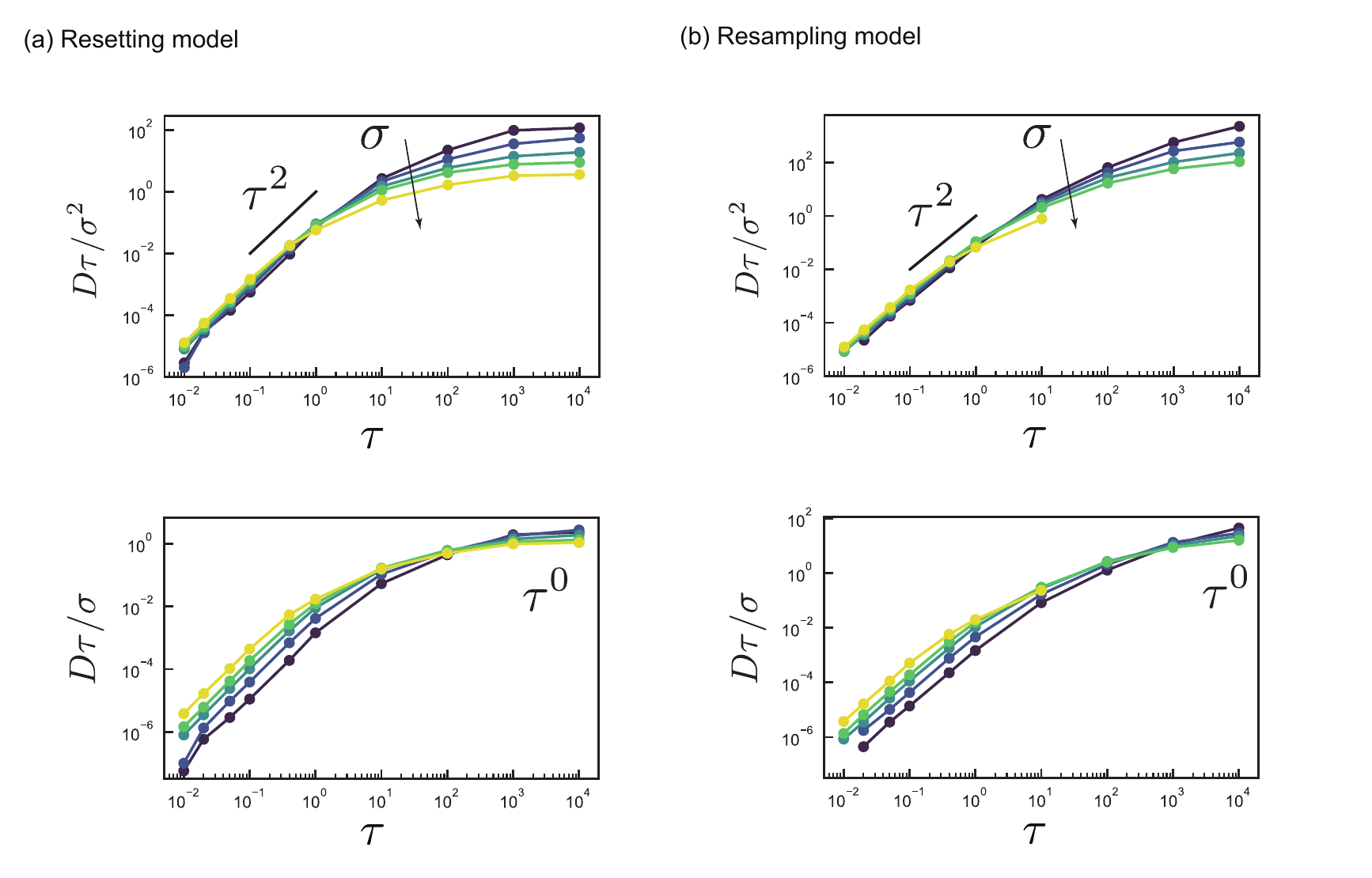}
 \end{center}
 \caption{\textbf{Scaling of $D$ with respect to $\sigma$ and $\tau$ in the small and large $\tau$ regime for the (a) resetting and (b) resampling models.} Data collapse in a plot of $D\tau/\sigma^2$ vs. $\tau$ demonstrates the scaling relation $D\propto\sigma^2\tau$ in the small $\tau$ regime. The solid line is a guide to the eye indicating the power law $\tau^2$. $\sigma \in [ 0.02, 0.05, 0.10,0.15, 0.30]$ (from dark color to light color).
 Data collapse in a plot of $D\tau/\sigma$ vs. $\tau$ demonstrates the scaling relation $D\propto\sigma/\tau$ in the large $\tau$ regime. 
 $\sigma \in [ 0.02, 0.05, 0.10,0.15, 0.30]$ (from dark color to light color).}
 \label{fig_SI_diffusion_constant}
\end{figure}

\ty{We show the scaling of the diffusion constant $D$ with respect to $\sigma$ and $\tau$ in the small and large $\tau$ regime for the resetting and resampling models in Fig.~\ref{fig_SI_diffusion_constant}. 
In both models, as in the persistent model, $D$ asymptotically approaches $D\sim\sigma^2\tau$ in the small $\tau$ regime and $D\sim\sigma/\tau$ in the large $\tau$ regime. We note that in some resampling model simulations where $\tau$ and $\sigma$ are near the extremes of model parameter values, numerical instabilities occur during the simulation and so we do not include those data points in the figures.}

\section{Analysis of trapped edges}
\mlm{We note that in the persistent model, a large number of very short edges appear when $\tau$ is large (Fig.~2(a-d)), suggesting that some edges might be trapped close to zero length.}

\mlm{To test whether these have an impact on the dynamics at large $\tau$ we focus on higher-order vertices, which are vertices where more than three cells meet, {\it i.e.} rosette structures. 
A CVM study by Yan {\it et al.} in the limit of zero fluctuations recently showed that rosette structures can rigidify the epithelial tissue~\cite{yanMulticellularRosettesDrive2019}.  We want to study whether the rigidification of the tissue driven by rosette structures slows down the dynamics in our model in the large-$\tau$ regime. }

However, unlike ref.2, by construction our model only contains 3-fold coordinated vertices.  Nevertheless, we hypothesize that in a dynamic simulation with finite fluctuations, vertices connected by very short interfaces restrict the dynamics in a manner similar to multi-fold coordinated vertices. Although higher-fold vertices are generically unstable in the fluid phase in CVMs with spatially homogeneous parameters~\cite{spencerVertexStabilityTopological2017}, some of us previously reported similar behavior in a 2D CVM with extra interfacial tensions between two cell types, where nearly-4-fold vertices (with very short edges) are stabilized at the heterotypic interface~\cite{sussmanSoftSharpInterfaces2018b}. Therefore it is not surprising that fluctuating heterotypic tensions could drive similar phenomena.

This is also consistent with our previous qualitative analysis of cellular structures:  Fig.~2 (b, c) shows that an increasing number of very short edges, highlighted by square symbols, is associated with rigidification in the large-$\tau$ regime.

To better understand how short edges affect the \ty{overall} dynamics in this model, we study their \ty{individual} dynamics. Specifically, we track edges, indexed by $i$, that reach the threshold $l_{\rm th} = 0.03$ for checking a T1 transition. At every subsequent timestep where the \ty{edge length} continuously remains below $l_{\rm th}$, we record the edge length $l^{i}(t_T)$, where $t_T$ is the time since the edge first crossed $l_{\rm th}$. 

\begin{figure}[h]
 \begin{center}
  \includegraphics[width=85mm]{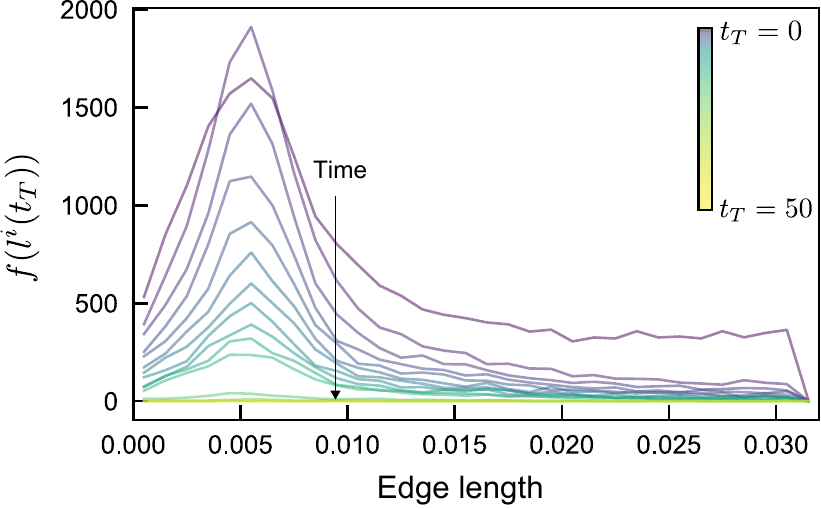}
 \end{center}
 \caption{\textbf{The time-evolution of the distribution $f(l^i(t_T))$ of the length $l^i(t_T)$ of the edges experiencing T1 events.} The data for $\tau=10$ and $\sigma=0.15$ is shown as an example. The color of the curves logarithmically maps the time $t_T$ ranging between $0$ and $50$ natural time unit. 
 }
 \label{app1}
\end{figure}

To quantify this behavior, we study histograms of the edge lengths $f(l^{i})$ for various values of \ty{this} trapping time $t_T$ (Fig.~\ref{app1}). 
For all but the longest timescales, there is a peak around $l \sim 0.005$, which is much smaller than the imposed T1 threshold $l_{\rm th}$, suggesting there is a population of edges where the dynamics drives them to remain very short. Such edges must remain short either because accepting a T1 transition increases the energy, and so T1 steps are rejected, or because they alternate between T1 events at every timestep.  In either case, the geometry and the tensions are such that it is energetically favorable for the edge to remain very short over multiple timesteps, resulting in a “trapped" short edge that functions very much like a multi-fold coordinated vertex. 

\begin{figure}[htb]
 \begin{center}
  \includegraphics[width=100mm]{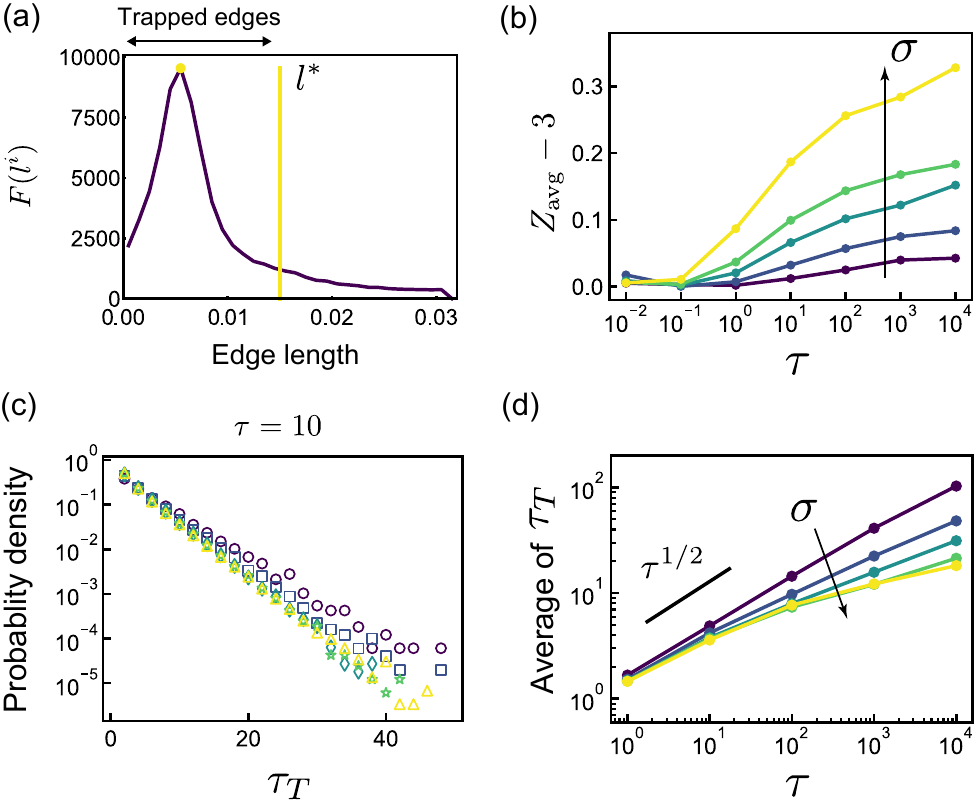}
 \end{center}
 \caption{\textbf{Properties of trapped edges.} (a) Time-integrated distribution of edge lengths, $F(l^i)$, defined in the main text. Trapped edges are defined as the population of edges in the peak, thresholded by the edge length by $l^*$ as indicated by the yellow vertical line. 
 (b) The average vertex coordination number per frame $Z_{\rm avg}$ subtracted by $3$ is plotted against $\tau$. In (b), we set $l^*=l_{\rm th}$ for some data points $(\tau,\sigma)=(0.01,0.1/0.15/0.3),(0.1,0.02)$, since the distributions $F(l^i)$ were too broad to determine the threshold $l^*$ for the trapped edges for these parameter values.
 (c) Semi-log plots of the probability distribution of the life time $\tau_T$ of the trapped edges for $\tau=10$. 
 (d) The average lifetime $\tau_T$ of the trapped edges is plotted against $\tau$. The solid line is a guide for eyes indicating the power law $\tau^{1/2}$. In (b-d), $\sigma \in [ 0.02, 0.05, 0.10,0.15, 0.30]$ (from dark color to light color).}
 \label{t1trap}
\end{figure}

Figure~\ref{t1trap}(a) shows an integral of these length histogram over all time windows, $F(l^i) = \sum_{t_T}{f(l^{i}(t_T))}$, which similarly exhibits a prominent peak $l \sim 0.05$, highlighting a characteristic length for edges that are trapped.  We randomly sampled 100 edges and 200 T1 events for each edge to plot Fig.~\ref{t1trap}(a). To formally define “trapped edges", we use this peak to define a new threshold lengthscale $l^*$ shown by the vertical line in  Fig.~\ref{t1trap}(a), that provides an upper bound to the length of the vast majority of trapped edges, see discussion in section \ref{appendix_def_threshold}.  This allows us to formally define all edges with length $l< l^*$ as trapped edges that may be functioning as “effective high-order vertices". We can also define a “trapped edge lifetime" $\tau_T$ corresponding to the number of natural time units where the edge continuously maintains a length less than $l^*$.

To quantify the density of these effective higher-order vertices, we calculate $Z_{\rm avg}$, which is the average vertex coordination number $Z = 2E/V$, where $E$ and $V$ are the number of edges and vertices per timestep \cite{yanMulticellularRosettesDrive2019} 
: if we have no trapped edges and only 3-fold vertices, $Z=3$.
We calculate $Z_{\rm avg}$ as $Z_{\rm avg}=2(E_0-T_{\rm avg})/(V_0-T_{\rm avg})$, where $T_{\rm avg}$ is the average number of trapped edges per timestep.
In Fig.~\ref{t1trap}(b), we plot $Z_{\rm avg}-3$ with respect to $\tau$ for different $\sigma$.  We find that $Z_{\rm avg}-3$ increases monotonically as $\tau$ and $\sigma$ increase.

\begin{figure}[h]
 \begin{center}
  \includegraphics[width=120mm]{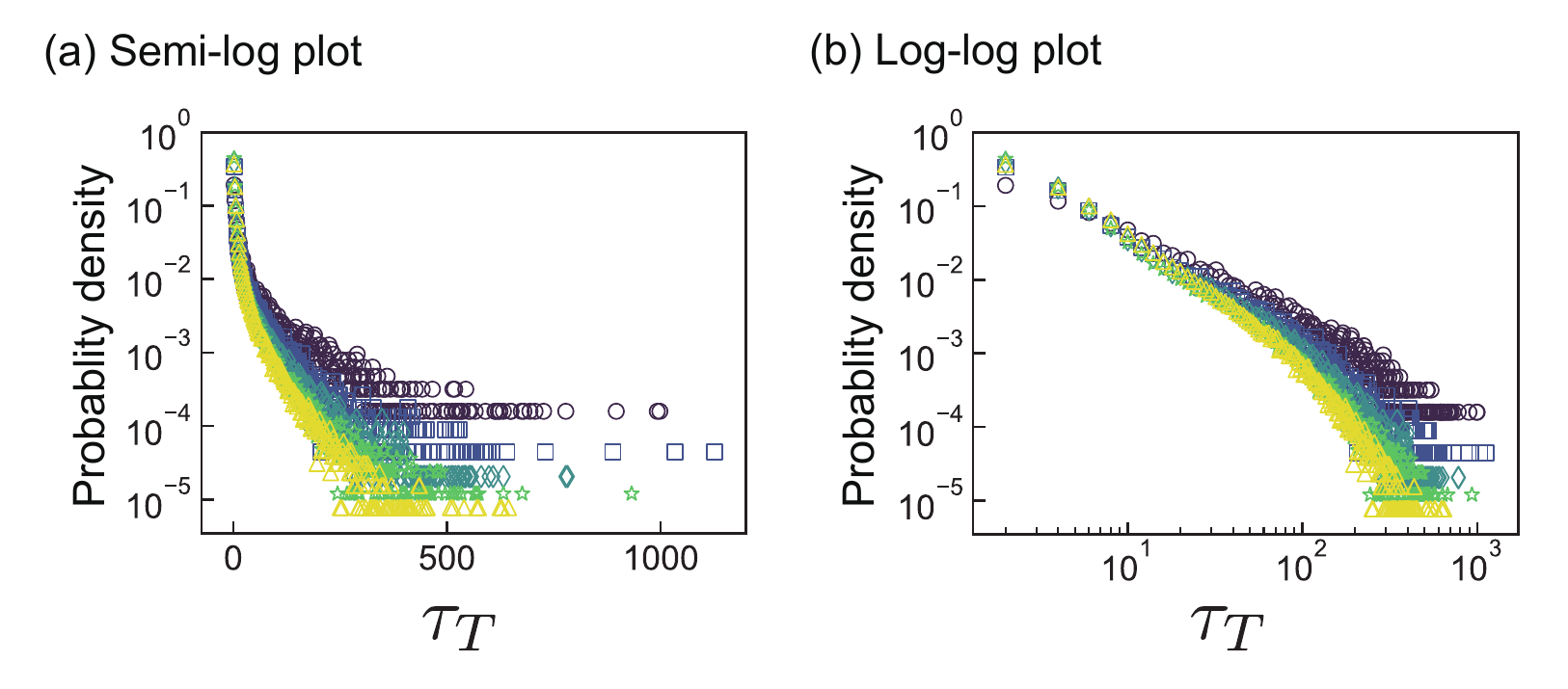}
 \end{center}
 \caption{\textbf{The probability distribution of the life time $\tau_T$ of the trapped edges.}
 (a) Semi-log and (b) log-log plots of the probability distribution $\tau_T$ for $\tau=1000$. $\sigma \in [ 0.02, 0.05, 0.10,0.15, 0.30]$ increase from dark color to light color (circle: $\sigma=0.02$, square: $\sigma=0.05$, diamond: $\sigma=0.10$, star: $\sigma=0.15$, triangle: $\sigma=0.30$).}
 \label{app2}
\end{figure}

Moreover, since our system is dynamic (unlike the system in ref.2), the persistence time of multi-fold coordinated vertices may be important.
Therefore, we also investigate the lifetime of trapped edges $\tau_T$, with normalized histograms shown in Fig.~\ref{t1trap}(c) and Fig.~\ref{app2}. The distribution is consistent with an exponential in the moderate $\tau$ regime ($\tau\sim 10$) as shown in Fig.~\ref{t1trap}(c), while it looks nearly power-law, with a large-scale cutoff, in the large-$\tau$ regime ($\tau\sim 1000$, see Fig.~\ref{app2}). Although the mechanisms driving these distributions remains unclear, we can nevertheless extract the average lifetime of trapped edges $\langle \tau_T\rangle$  as a function of model parameters, shown in Fig.~\ref{t1trap}(d). The average lifetime of trapped edges increases dramatically with increasing $\tau$, and also increases slightly with decreasing $\sigma$.  Taken together, these results suggest that there is a systematic increase in the fraction and persistence of effectively multi-fold coordinated vertices at large $\tau$, which, in the absence of other effects, should tend to rigidify the system.

One obvious question, especially given the important role of effective multi-fold coordinated vertices, is whether our results depend strongly on our choice of how to resample the stress in the newly created edges after a T1 swap.  The “persistent" model we have considered so far gives the new edge after a T1 swap the same tension as the old edge, which will clearly favor trapped edges where the tension is larger and contractile.  Therefore, we also investigate more democratic ways of sampling tensions in the new T1 edge, illustrated schematically in Fig.~1 (b) and (c), which we term “resetting" and “resampling" models. 

 Figure~6 shows that, as expected, resetting and resampling models generate the same diffusion constants as the persistent models in the small-$\tau$ regimes, consistent with the hypothesis that fluctuation-driven diffusion, which should be the same in all models, dominates at low $\tau$. In addition, there is still non-monotonic behavior in all three models, with the diffusion constant decreasing at large $\tau$\textcolor{black}{, while there is an obvious increase in the diffusion constant at larger $\tau$ in the resetting and resampling models compared to the persistent model.}
 
\begin{figure}[t]
 \begin{center}
  \includegraphics[width=178mm]{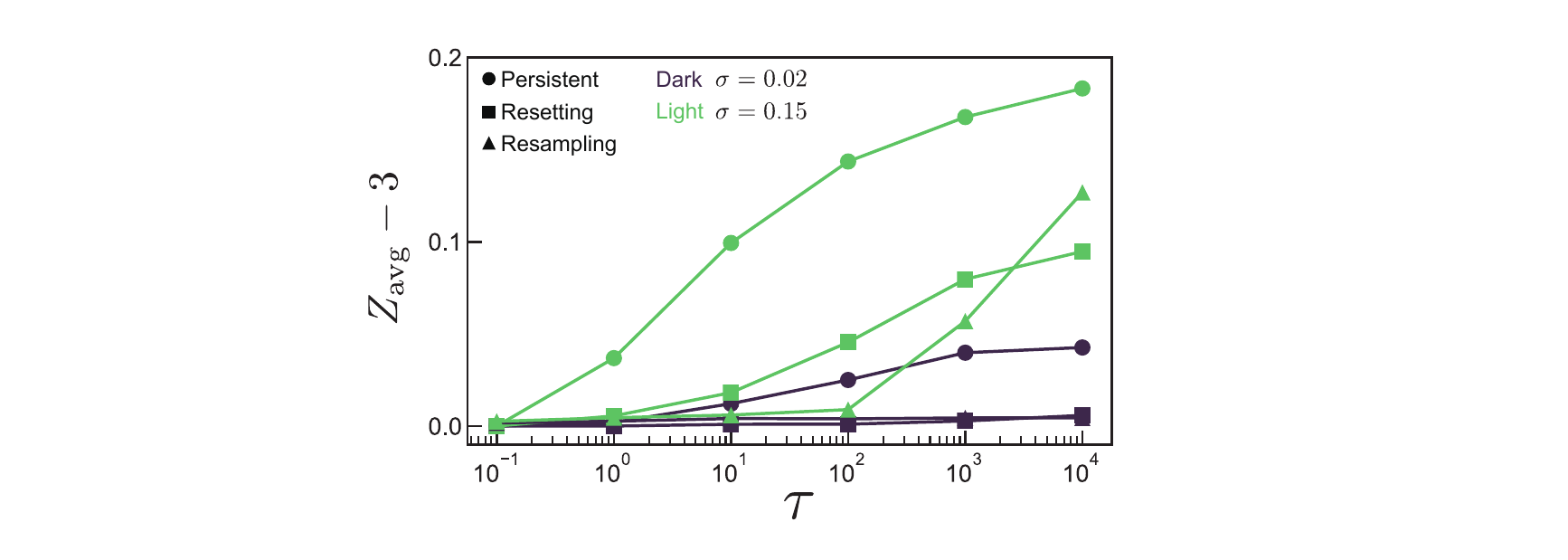}
 \end{center}
 \caption{\textbf{Comparison $Z_{\rm avg}-3$ between three models}: the persistent model (circle), the resetting model (square), the resampling model (triangle). $Z_{\rm avg}-3$ vs. $\tau$ for the three models, respectively. Dark and light markers represent the data with $\sigma=0.02$ and $\sigma=0.15$, respectively.
The distributions $F(l^i)$ were too broad to determine the threshold $l^*$ for the trapping edges in the following data points: $(\tau,\sigma)=(0.1,0.02)$ in the persistent model, $(\tau,\sigma)=(0.1,0.02/0.15),(1,0.02)$ in the resetting model, $(\tau,\sigma)=(0.1,0.02/0.15),$ $(1,0.02/0.15),$ $(10,0.02/0.15),$ $(100,0.02),$ $(1000,0.02),$ $(10000,0.02)$ in the resampling model.
We hence set $l^*=l_{\rm th}$ for these data points.
 }
 \label{3model_result}
\end{figure}

\begin{figure}[h]
 \begin{center}
  \includegraphics[width=120mm]{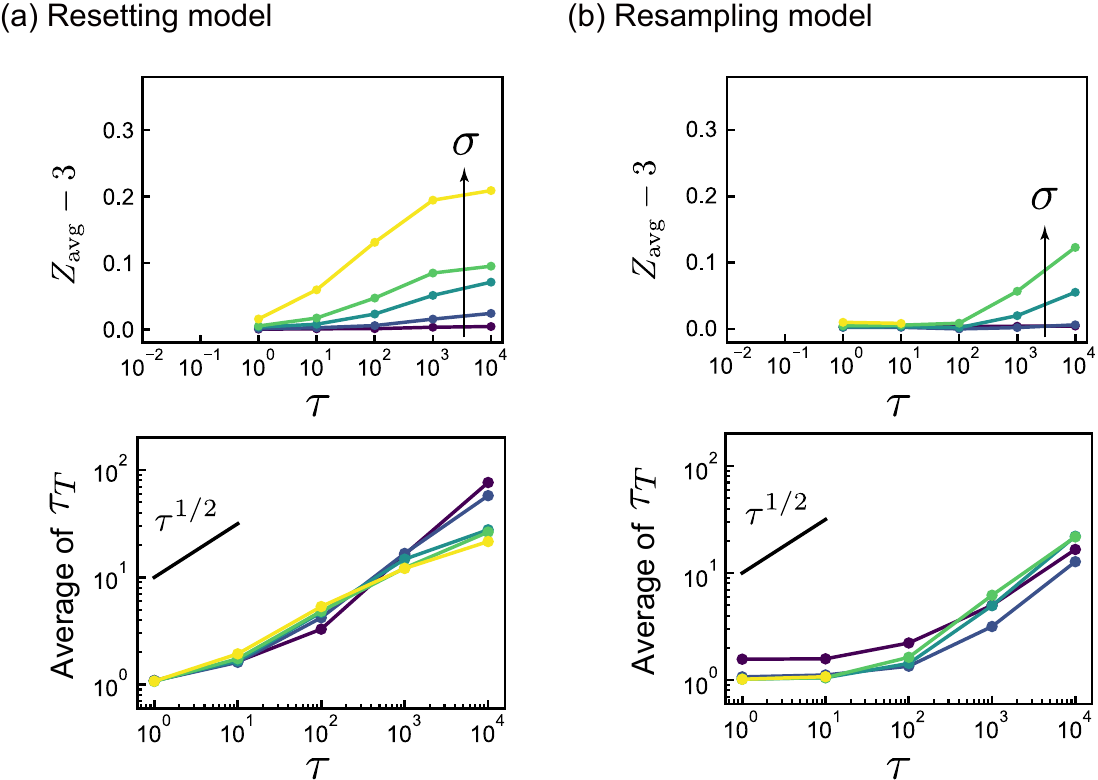}
 \end{center}
 \caption{\textbf{Properties of trapped edges for the resetting and resampling models.} In the top panels of (a) resetting model and (b) resampling model, the average vertex coordination number per frame $Z_{\rm avg}$ subtracted by $3$ is plotted against $\tau$. In the bottom panels, the average lifetime $\tau_T$ of the trapped edges is plotted against $\tau$. 
 The distributions $F(l^i)$ were too broad to determine the threshold $l^*$ for the trapping edges in the following data points: $(\tau,\sigma)=(1,0.02)$ in the resetting model and $(\tau,\sigma)=(1,0.02/0.05/0.1/0.15/0.3),(10,0.02/0.05/0.1/0.15),(100,0.02),(1000,0.02),(10000,0.02)$ in the resampling model.
 The solid line is a guide for eyes indicating the power law $\tau^{1/2}$. In (a) and (b), $\sigma \in [ 0.02, 0.05, 0.10,0.15, 0.30]$ (from dark color to light color).}
 \label{app6}
\end{figure}

  \textcolor{black}{In order to see if multi-fold coordinated vertices play an important role in rigidification in the large $\tau$ regime of the resetting and resampling models, we measured the fraction of trapped edges and the trapping time for both models. In general, we follow the same procedure outlined above for the persistent model. We note that in some resampling model simulations where $\tau$ and $\sigma$ are near the extremes of model parameter values, numerical instabilities occur during the simulation and so we do not include those data points in the figures. As shown in Figs.~\ref{3model_result} and \ref{app6}, especially in the resampling model, there was not a significant increase in the number of trapped edges, and again only a modest increase in the trapping time.  Given that the resampling model also exhibits a nonmonotonic curve for the diffusion constant with increasing $\tau$, this suggests that, at least in the resampling case, the trapped edges are not responsible for the decrease in diffusion observed at large $\tau$.
 We conclude that rigidification driven by multi-fold coordinated vertices is a potential mechanism contributing to reduced diffusion in some models, but clearly not sufficient to explain the non-monotonoicity in all the models we studied. An interesting question for future work remains whether and how short trapped edges are contributing to the rigidification in the persistent model, where they are common.}

\subsection{Definition of the threshold $l^*$ of the trapped edges\label{appendix_def_threshold}}

We first detected a maximum peak at $l=l_{\rm max}$ in the time-integrated distribution $F(l^i)$ as indicated by a yellow circle marker in Fig.~\ref{t1trap}(a). 
We next subtracted the minimum frequency in the range $l_{\rm max}\leq l\leq l_{\rm th}$, as the background, from the time-integrated distribution. 
Using this background-subtracted distribution, we finally determined the threshold $l^*$ as the minimum edge length at which the frequency is below $10\%$ of the maximum frequency at $l=l_{\rm max}$. In Fig.~\ref{app1}, we show an example of histograms of the edge lengths $f(l^{i})$ for various values of the trapping time $t_T$.

\bibliography{rsc} 